\documentclass{aa}  

\usepackage{graphicx}	
\usepackage{txfonts}
\usepackage{hyperref}
\usepackage{placeins}

\newcommand{\chandra}{\textit{Chandra}}	
\newcommand{\xmm}{\textit{XMM-Newton}}
\newcommand{\rosat}{\textit{ROSAT}}

\begin{document}

   \title{X-ray investigation of the remarkable galaxy group Nest200047}
   \subtitle{}
   \titlerunning{Nest200047}
   \authorrunning{Anwesh Majumder et al.}

   \author{Anwesh Majumder
          \inst{1,2}\fnmsep\thanks{Email: a.majumder@sron.nl},
          A. Simionescu
          \inst{2,3,4}\fnmsep\thanks{Email: a.simionescu@sron.nl},
          T. Pl\v{s}ek
          \inst{5},
          M. Brienza
          \inst{6},
          E. Churazov
          \inst{7,8},
          I. Khabibullin
          \inst{9,7,8},
          F. Gastaldello
          \inst{10},
          A. Botteon
          \inst{6},
          H. R\"ottgering
          \inst{3},
          M. Br\"uggen
          \inst{11},
          N. Lyskova
          \inst{8,12},
          K. Rajpurohit
          \inst{13},
          R.A. Sunyaev
          \inst{8,7}
          \and
          M.W. Wise
          \inst{2,1}
          }

   \institute{Astronomical Institute `Anton Pannekoek’, University of Amsterdam, Science Park 904, 1098 XH Amsterdam, The Netherlands
        \and
             SRON, Netherlands Institute for Space Research, Niels Bohrweg 4, 2333 CA Leiden, The Netherlands 
        \and
             Leiden Observatory, Leiden University, PO Box 9513, 2300 RA Leiden, The Netherlands
        \and
             Kavli Institute for the Physics and Mathematics of the Universe, The University of Tokyo, Kashiwa, Chiba 277-8583, Japan
         \and
             Department of Theoretical Physics and Astrophysics, Faculty of Science, Masaryk University, Kotl\'a\v{r}sk\'a 2, 602 00 Brno, Czech Republic
        \and
            Istituto Nazionale di Astrofisica (INAF) - Istituto di Radioastronomia (IRA), via P. Gobetti 101, 40129 Bologna, Italy
         \and
            Max Planck Institute for Astrophysics, Karl-Schwarzschild-Str. 1, D-85741 Garching, Germany
        \and
            Space Research Institute (IKI), Profsoyuznaya 84/32, Moscow 117997, Russia
        \and
            Universit\"ats-Sternwarte, Fakult\"at f\"ur Physik, Ludwig-Maximilians-Universit\"at M\"unchen, Scheinerstr.1, 81679 M\"unchen, Germany
        \and
            INAF/IASF-Milano, Via A. Corti 12, I-20133 Milano, Italy
        \and
            University of Hamburg, Hamburger Sternwarte, Gojenbergsweg 112, 21029 Hamburg, Germany
        \and
            Astro Space Center of P.N. Lebedev Physical Institute of the RAS, Profsoyuznaja 84/32, 117997 Moscow, Russia
        \and
            Center for Astrophysics | Harvard \& Smithsonian, 60 Garden Street, Cambridge, MA 02138, USA}

   \date{Received XXX / Accepted XXX}
 
  \abstract
   {Galaxy groups are more susceptible to feedback from the central active galactic nuclei (AGNs) due to their lower gravitational binding energy compared to clusters. This makes them ideal laboratories to study feedback effects on the overall energy and baryonic mass budget.}
   {We study the LOFAR-detected galaxy group Nest200047, where there is clear evidence of multiple generations of radio lobes from the AGN. Using 140 ks \textit{Chandra} and 25 ks \textit{XMM-Newton} data, we investigated thermodynamic properties of the intragroup medium, including any excess energy due to the central AGN. We also investigated the X-ray properties of the central black hole and constrained the $2-10$ keV X-ray flux.}
   {We used spectral analysis techniques to measure various thermodynamic profiles across the whole field of view. We also used both imaging and spectral analysis to detect and estimated the energy deposited by potential shocks and cavities. Due to the faint emission from the object beyond the core, various background effects were considered.}
   {Nest200047 has significant excess entropy, and the AGN likely contributes to a part of it. There is an excess energy of $(5-6.5) \times 10^{60}$ erg within 400 kpc, exceeding the binding energy. The pressure profile indicates that gas is likely being ejected from the system, resulting in a baryon fraction of $\sim4\%$ inside $r_{500}$. From scaling relations, we estimated a black hole mass of $(1-4)\times 10^9 M_{\odot}$. An upper limit of $2.1 \times 10^{40}$ erg s$^{-1}$ was derived on the black hole bolometric luminosity, which is $\sim$2.5\% of the Bondi accretion power.}
   {Nest200047 is likely part of a class of over-heated galaxy groups, such as ESO 3060170, AWM 4, and AWM 5. Such excessive heating may lead to high quenching of star formation. Moreover, the faint X-ray nuclear emission in Nest is likely due to the accretion energy being converted into jets rather than radiation.}


   \maketitle

\section{Introduction}

\begin{figure*}
\includegraphics[width=\textwidth]{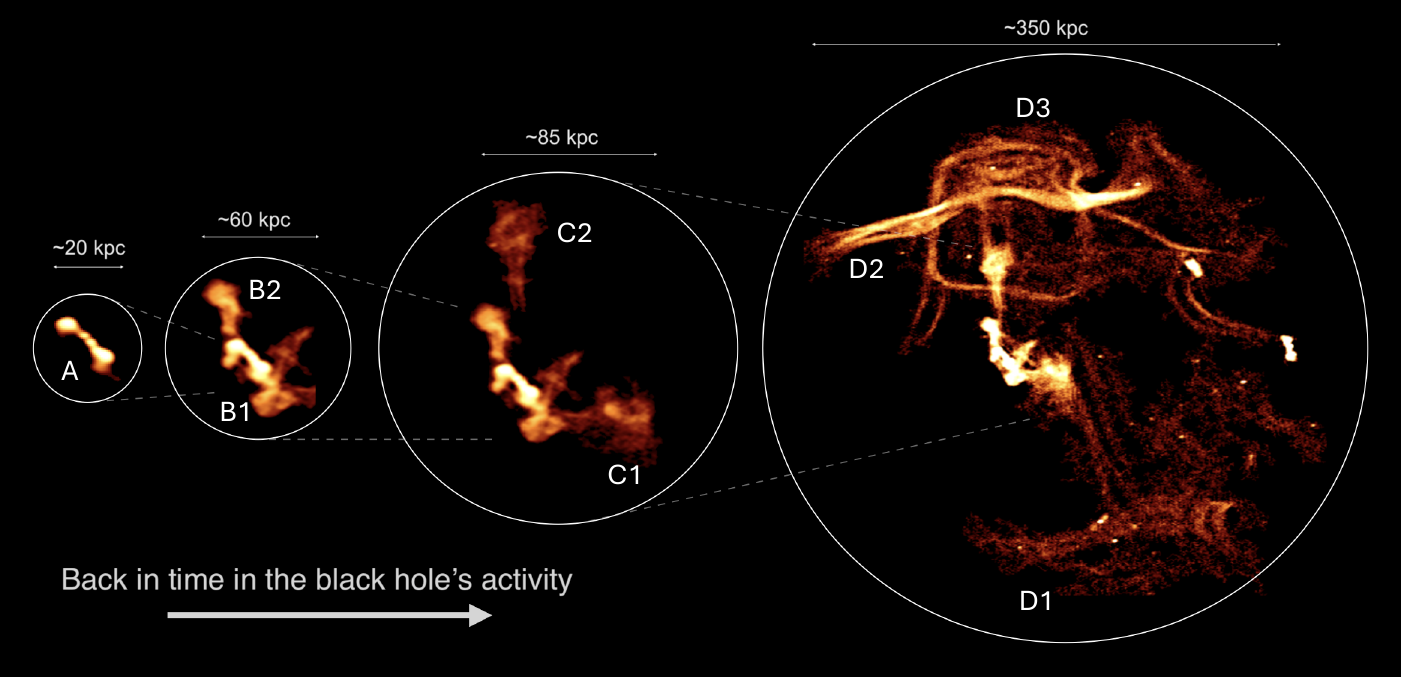}
\caption{\small{
LOFAR image of the galaxy group Nest200047, focusing on various spatial scales to highlight the four consecutive AGN outbursts detected in this system \citep{bri21}.}}
\label{NEST_LOFAR}
\end{figure*}

Radio-mode feedback from active galactic nuclei (AGNs) plays a key role in regulating the cooling and star formation in a broad range of halo masses \citep{chu00,bir04,bes06,wer19}. The jets from the central AGN often create observable structures, such as shocks, that transfer energy to the ambient medium and X-ray cavities by expelling gas from the core (e.g., \citealt{beg89,mcn01,fab06,wis07,sim09}). Although our understanding of such mechanisms has improved over the years (for reviews, see \citealt{mcn07,sim19}), the effect of multiple generations of AGN outbursts on the ambient medium remains an interesting problem. Multi-generational outbursts are expected from numerical simulations (e.g., \citealt{rosa16,deg17}), and their signatures have been observed in the radio band as multiple pairs of radio lobes interacting with the ambient medium (see \citealt{mor17,mor24} for reviews). Such evidence of recurrent outbursts has also been seen in X-rays. Although more recent outbursts are often associated with cavities, the older outbursts have different evolutionary paths that sometimes create X-ray bright arms of low-entropy gas (M87; \citealt{for07,wer10}) or density and temperature ripples (Cygnus A; \citealt{maj24}). Extending such studies to a larger number of multiple-outburst systems is important for further pinpointing the details of AGN feedback evolution.

While known examples of multiple generations of outbursts are mostly limited to clusters (e.g., \citealt{vant14,bia21}), the famous example of NGC 5813 and NGC 5044 \citep{ran11,sch21} shows that such an evolutionary history is possible for galaxy groups as well. Galaxy groups lie in a critical mass regime ($10^{13}-10^{14} M_{\odot}$), where the hot gas is still sufficiently bright to be observable in X-rays but where AGN feedback may more closely resemble the mechanism needed to quench star formation in individual galaxy haloes. In particular, the energy supplied by the central AGN can exceed the gravitational binding energy of halo gas particles, which is not the case in more massive galaxy clusters \citep{eck21,ayr23}. Therefore, further exploring the impact of multi-generation AGN feedback on the intragroup medium (IGrM)  can lead to a better understanding of their impact on the evolutionary history of massive galaxies. 

The Low-Frequency Array (LOFAR) Two-Meter Sky Survey (LoTSS; \citealt{shi19}) and the extended ROentgen Survey with an Imaging Telescope Array (eROSITA)  X-ray telescope \citep{pred21} on board the Spektrum
Roentgen Gamma (SRG) orbital observatory \citep{sunyaev21} have recently revealed a truly spectacular example of recurring AGN feedback in a largely unexplored, nearby  ($z = 0.01795 \pm 0.00015$; \citealt{huc12}) galaxy group called Nest200047 (2MASS survey; \citealt{tul15}). The radio and X-ray properties of the object from those investigations were first reported by \cite{bri21}. In the radio band, two freshly injected radio lobes, each roughly extending $30$ arcsec ($11$ kpc in projection) in length from the group center, are being inflated by the central AGN associated with the brightest group galaxy, MCG+05-10-007 (A lobes in Figure \ref{NEST_LOFAR}). Farther away, two more radio bubbles (B1 and B2) extend out to radii of 100 arcsec ($37$ kpc) from the optical centroid of the host galaxy. A third pair of radio lobes (C1 and C2) can be seen between $100-275$ arcsec ($37-100$ kpc) from the AGN, while farther still, the lobes D1 and D3 extend out to a $10$ arcmin radius ($220$ kpc), rivaling the large radio structures found in more massive galaxy clusters \citep{lane04,mcn05,wis07}. The outermost lobes exhibit a striking array of complex filamentary radio structures (D2), likely preserved by magnetic fields, showing sharp bends and double strands of very narrow emission. They may represent a rare case of an AGN bubble in its very late stages of evolution, as it is being deformed and shredded by instabilities and possibly in the process of being mixed with the IGrM on small scales. No other radio galaxy in the current LoTSS database shares a similar morphology. The shallow (645 s) eROSITA scan of the object \citep{bri21}, on the other hand, revealed edge-brightening around the radio bubbles that were classified as cavities with powers upwards of $10^{42}$ erg s$^{-1}$. The object has a temperature of $\sim2$ keV within 13 arcmin and a $0.5-2.0$ keV luminosity of $(5-10) \times 10^{42}$ erg s$^{-1}$ within 30 arcmin. More recently, \cite{bri25} performed a more detailed spectral analysis over a broad radio band (53-1518 MHz) using the upgraded Giant Metrewave Radio Telescope (uGMRT; \citealt{gup17}), MeerKAT \citep{jon16}, and the Karl G. Jansky Very Large Array (VLA; \citealt{per11}). This recent work estimated the duration of the jet activity for different bubbles to be between 50-100 Myr.

In this study, we report the results of the first pointed X-ray observations performed for this galaxy group. We present the X-ray morphology and thermodynamic profiles of the object and estimate energy injection by the radio lobes using \textit{Chandra} and \textit{XMM-Newton} data. The high spatial resolution of \textit{Chandra} allows us to probe any deviation from the expected smooth thermodynamic profiles under hydrostatic equilibrium, while XMM's sensitivity at low energy allows us to probe the outskirts more effectively. Since the object is nearby, \textit{Chandra}'s spatial resolution also allows us to put constraints on the black hole mass and X-ray properties of the central AGN. 

\begin{table}
    \centering
\caption{{\chandra} and {\xmm} observation log.}
\label{tab:exposure_info}
    \begin{tabular}{crcr}
        \hline
        \hline
         Telescope&  \multicolumn{1}{c}{ObsID}&  Instrument& \multicolumn{1}{c}{Exposure (ks)}\\
         \hline
         &   &  EMOS1& 26.8\\
         \xmm&  0883620101&  EMOS2& 27.8\\
         &  &  EPN& 9.8\\
         \hline
         &  26949&  & 14.3\\
         &  27221&  & 12.3\\
         &  27222&  & 14.4\\
         &  27223&  & 14.0\\
         \chandra&  27224&  ACIS-I& 14.0\\
         &  27571&  & 16.4\\
 & 27582& &11.0\\
 & 27598& &14.7\\
 & 27607& &14.3\\
 & 27608& &14.6\\
 \hline
 & & Total&140\\
\hline
\hline
    \end{tabular}  
    \tablefoot{The exposure times noted are after flare cleaning.}
\end{table}

We begin in \S\ref{Data} by presenting the observations and the data reduction methods applied in this work. In \S\ref{lss}, we describe the large-scale features in the X-ray data and highlight any hints of AGN feedback. A detailed description of the spectral analysis method is discussed in \S\ref{spec_analysis}, with more details in \S\ref{CXB_flux} and \S\ref{soft_proton}. All thermodynamic profiles are then shown and discussed in \S\ref{profiles}. Possible deviations from smooth density profiles are discussed in \S\ref{dens_break} and then followed by an estimation of the energetics of a tentatively detected cavity in \S\ref{cavity}. The mass estimate and X-ray properties of the central supermassive black hole follow in \S\ref{SMBH}. We conclude with a summary and conclusion of our results in \S\ref{discussion}.

A $\Lambda$CDM cosmology has been assumed in this work with $H_0 = 70$ km s$^{-1}$ Mpc$^{-1}$, $\Omega_{\Lambda} = 0.7$, and $\Omega_m = 0.3$. The corresponding linear scale at the redshift of Nest200047 is 21.9 kpc per arcmin. All errors reported are of $1\sigma$ significance.

\section{Observations and data reduction} \label{Data}

\subsection{\xmm}

Nest200047 (hereafter Nest) was observed by XMM-Newton on 13-09-2021 for a total exposure of 46 ks. We analyzed the exposure (ObsID: 0883620101) with Science Analysis Software (SAS) v21.0.0 and the latest calibration files. The observation was heavily affected by soft proton flares. We reprocessed the MOS and PN event files from the Observation Data Files (ODF) with SAS tools \texttt{emproc} and \texttt{epproc}, respectively. Furthermore, we created a separate out-of-time (OoT) event file for the PN observation with \texttt{epproc}. The MOS event files were then checked for anomalous soft X-ray noise \citep{kun08,stuh08} with the SAS tool \texttt{emanom}. The tool flagged CCD 4 of MOS1 as ``bad" and CCD 5 of MOS2 as ``intermediate." The MOS1 CCD 4 was henceforth removed from all subsequent analysis, while CCD 5 of MOS2 was kept after checking the spectrum from part of this chip visually. Despite this cleaning, we still found anomalous soft X-ray noise below 1 keV in MOS1 spectra from the central CCD upon visual inspection. We, therefore, discarded the MOS1 spectrum from the central CCD below 1 keV for spectral analysis (see \S\ref{soft_noise} for more discussion). 

To reduce the heavy soft proton contamination, we filtered all event files as follows: The MOS events were binned in the energy range of $10-12$ keV in 100 s intervals. We did the same with the PN events but in the energy range of $10-14$ keV. We then fitted each count rate histogram with a Gaussian function to determine the mean ($\mu$) and the standard deviation ($\sigma$). All events with a count rate greater than $1\sigma$ from the mean were removed from all subsequent analysis.  We then focused on removing extremely fast flares by binning the events in the aforementioned energy ranges in 20 s intervals and repeated the Gaussian fits. All events with a count rate greater than $2\sigma$ from the mean were removed from all subsequent analysis as well. We report the clean exposure times from each instrument in Table \ref{tab:exposure_info}. For PN, we applied the same Good Time Intervals (GTIs) to the OoT event file as to the normal observation.

For suitable MOS particle background event lists, we used the integrated filter wheel closed (FWC) data\footnote{\href{https://www.cosmos.esa.int/web/xmm-newton/filter-closed}{https://www.cosmos.esa.int/web/xmm-newton/filter-closed}} (Rev. 230 - Rev. 4027).  Since there is no suitable out of field of view (FOV) region for PN \citep{zha20,mar21}, we chose a PN FWC observation (Event list: 3927\_0810811601\_EPN\_S017) that was performed soon before the Nest observation. The FWC files of each instrument were then reprojected onto the same sky coordinates as the Nest observation using the SAS task \texttt{evproject}.

All FOV events in both the observation and the FWC files were screened with the conditions \texttt{PATTERN<=12} and \texttt{(FLAG \& 0x766ba000)==0} for MOS, while the condition  \texttt{PATTERN<=4} and \texttt{FLAG==0} was chosen for PN. Separate out of FOV event files were also created with the condition \texttt{\#XMMEA\_16} for imaging analysis. This flag screens bad event grades but keeps valid out of field of view events.

\begin{figure*}
    \centering
    \includegraphics[width=\columnwidth, keepaspectratio]{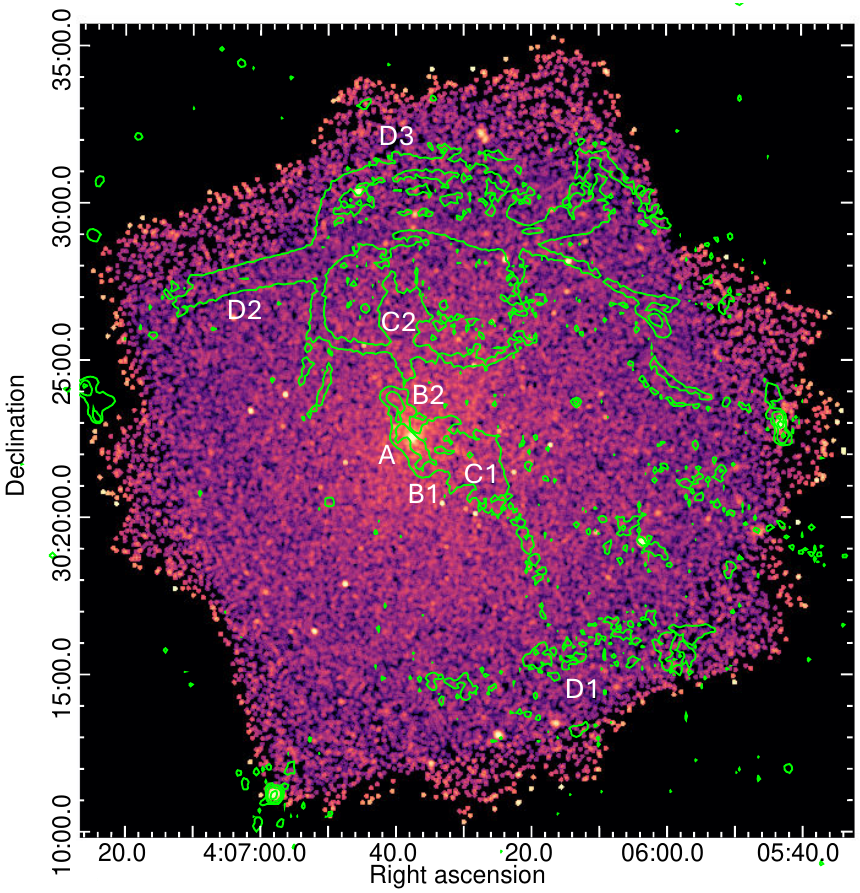}
     \includegraphics[width=\columnwidth,  height=3.59in]{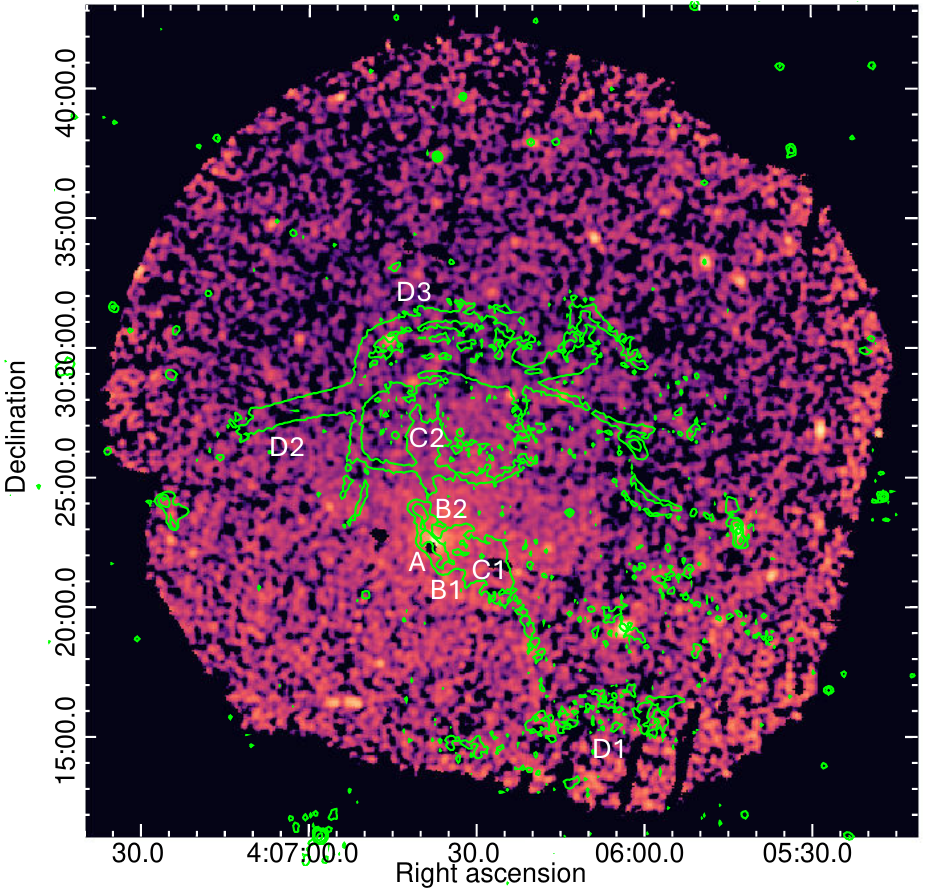}
    \caption{\small{\emph{Left:} {\chandra} exposure-corrected, background (NXB + X-ray foreground and background) subtracted image of Nest. \emph{Right:} {\xmm} exposure-corrected, background (NXB + X-ray foreground and background + soft proton) subtracted image of Nest. Both of these images were created in the $0.6-4.0$ keV energy band. The images have been Gaussian-smoothed with a $\sigma=3$ pixels for better visualization. The LOFAR 144 MHz radio contours from \cite{bri21} are shown in green. The contour levels used were 3, 10, 25, and 60 $\times$ $\sigma$, where $\sigma = 154 \mu$Jy/beam and the beam size is $8.6'' \times 4.3''$. The different lobes have been marked on each figure.}}
    \label{fig:fov}
\end{figure*}

\subsection{\chandra}

The galaxy group Nest has been observed for a total of $\sim$$150$ ks with {\chandra} from 21-11-2022 to 18-12-2022. We reprocessed and analyzed all observations in the {\chandra} data archive using \texttt{CIAO 4.15} \citep{fru06} and \texttt{CALDB 4.10.7}. We used the \texttt{CIAO} tool \texttt{chandra\_repro} to create level 2 event files along with the condition \texttt{check\_vf\_pha = yes} for additional cleaning of particle background in VFAINT mode observations. All reprocessed event files were checked for flares using the same procedure applied for the XMM observations in the $9-12$ keV band. No strong background flares were found for any Observation ID. We report all cleaned exposure times in Table \ref{tab:exposure_info}. We furthermore excluded all off-axis ACIS-S chips and used only the ACIS-I chips that have lower background\footnote{\href{https://cxc.harvard.edu/proposer/POG/html/chap6.html}{https://cxc.harvard.edu/proposer/POG/html/chap6.html}}.

The particle background event list was obtained by using the ``stowed" ACIS event files from \texttt{CALDB}. The ``stowed" particle background was measured with the instrument unexposed to the sky. We created tailored particle background event files for each ObsID by applying appropriate gain corrections followed by their reprojection to the same sky plane as the observation. We filtered these particle background events further by selecting \texttt{status=0} events for VFAINT mode observations. For the rest of this paper, we will refer to the particle background as the non-X-ray background (NXB).

\begin{table}
    \centering
\caption{Unabsorbed galactic emission components from the RASS spectrum.}
\label{tab:galactic_fits}
    \begin{tabular}{ccc}
    \hline
    \hline
         Components&\  Flux ($0.1-2.4$ keV)& kT\\
         &  (photons s$^{-1}$ cm$^{-2}$)& (keV)\\
         \hline
         Galactic halo&  $6.8^{+1.0}_{-0.8}$ $\times 10^{-6}$& $0.217 \pm 0.013$ \\
         Local hot bubble&  ($4.42 \pm 0.11$) $\times 10^{-6}$& ($7.3 \pm 0.2$) $\times 10^{-2}$\\
    \hline
    \hline
    \end{tabular}
    \tablefoot{The above components were constrained by the RASS spectrum in an annular region between $1.5R_{200}$ and $1.5R_{200}+1^{\circ}$. The normalizations have been scaled to 1 arcmin$^2$ area.}
\end{table}

\subsection{X-ray background} \label{sky_bkg}

We modeled the Galactic X-ray emission and the cosmic X-ray background (CXB) using a spectrum from the {\rosat} all-sky survey (RASS). This spectrum was generated by the RASS X-ray background tool \citep{sab19}\footnote{\href{https://heasarc.gsfc.nasa.gov/cgi-bin/Tools/xraybg/xraybg.pl}{https://heasarc.gsfc.nasa.gov/cgi-bin/Tools/xraybg/xraybg.pl}}. We extracted the spectrum from a region far enough away from the object to ensure maximum background emission and minimum contamination from the object. We achieved this by choosing an annulus centered on the Nest galaxy group with an inner radius of $1.5R_{200}$ and outer radius of $1.5R_{200} + 1^{\circ}$, where $R_{200}$ was calculated for the Nest galaxy group. Using the upper limit of $M_{500} = 7\times10^{13}M_{\odot}$ reported by \cite{bri21}, we obtain a $R_{200}$ value of $\sim$$940$ kpc which is equivalent to $\sim$$0.7$ degree. The total number of spectral counts obtained in the $0.1-2.4$ keV band is $4450$.

Using \texttt{SPEX v3.07.03}\footnote{\href{https://spex-xray.github.io/spex-help/index.html}{https://spex-xray.github.io/spex-help/index.html}} \citep{kaa96,kaa18}, we fit a \texttt{hot(cie + pow) + cie} model to this spectrum. The components represent the absorbed galactic halo, absorbed CXB, and the unabsorbed local hot bubble, respectively. The temperature of the \texttt{hot} component was fixed to $10^{-6}$ keV for neutral absorption. We set the abundance value of each \texttt{cie} component to 1.0 with respect to the reference proto-solar abundance model of \cite{lod09}. The hydrogen column density in the field of view of our object was calculated using the method proposed by \cite{wil13}\footnote{\href{https://www.swift.ac.uk/analysis/nhtot/}{https://www.swift.ac.uk/analysis/nhtot/}}. The weighted hydrogen column density was found to be $n_H = 1.82 \times 10^{21}$ atoms cm$^{-2}$.  The CXB normalization was fixed to $8.88 \times 10^{-7}$ photons/s/cm$^2$/keV (for 1 arcmin$^2$ area at 1 keV, see \citealt{sno08}), while the index was fixed to 1.46. We report the resulting fit parameters in Table \ref{tab:galactic_fits}. We used C-statistic for minimization during fitting \citep{cash79,kaa17}. The \texttt{cstat} value after the fit was $11$ with the expected cstat value being $7 \pm 3$ (see \citealt{kaa17} for an explanation of expected C-statistic).

Since {\chandra} and {\xmm} resolve more point-sources than {\rosat}, the flux of unresolved point-sources for these two X-ray telescopes needs to be calculated from the logN-logS relation, assumed to follow that from the {\chandra} Deep Field South \citep{leh12}. The details of this calculation are discussed in Appendix \ref{CXB_flux}.

\subsection{Image processing}

The {\xmm} counts images for all the EPIC (European Photon Imaging Camera) instruments were created in the energy range $0.6-4.0$ keV and binned to 4-arcsec by setting the \texttt{ximagebinsize} and \texttt{yimagebinsize} parameters to $80$ in the SAS task \texttt{evselect}. A PN OoT image was also created in the same way and was subtracted from the PN observation image after scaling by a factor of $0.023$\footnote{\href{https://xmm-tools.cosmos.esa.int/external/xmm_user_support/documentation/uhb/epicoot.html}{https://xmm-tools.cosmos.esa.int/external/xmm\_user\_support/\\documentation/uhb/epicoot.html}}, appropriate for the extended full-frame mode. Exposure maps were then obtained for each instrument in the same energy band.  The standard XMM exposure maps for each instrument (in units of seconds) were multiplied with the on-axis effective area before combining to take into consideration different effective areas between the instruments. Finally, counts images and the altered exposure maps from all instruments were combined and used for all subsequent imaging analysis. We detected point sources with the help of the \texttt{CIAO} tool \texttt{wavdetect} and visually confirmed them before removing them from any spectral or imaging analysis.

Particle background images were created from the FWC event files using the same binning and energy range as the observation images. The MOS background images were normalized using the unexposed area of the instrument as described by \cite{kun08}. On the other hand, the PN background image was normalized using the whole field of view counts in the $10-14$ keV band. These particle background images were then combined and used to obtain the total NXB image. The exposure-corrected, background-subtracted images for the full FOV of {\xmm} are shown in the right panel of Figure \ref{fig:fov}.

For {\chandra}, all level 2 event files were first reprojected to a common tangent point. Count images and exposure maps were created in the $0.6-4.0$ keV band for each reprojected event file and then combined to produce the total counts and exposure map. Point sources were again detected with the help of \texttt{wavdetect}. The point sources were removed in all spectral analysis. We also removed them from all image analysis in \S\ref{profiles} to \ref{SMBH}. 

Background images for individual reprojected stowed event files were created in the same energy band as observation images and were scaled to match the $9-12$ keV band emission. Similar to our XMM background image, these ``stowed" event file images were combined and used to obtain total NXB image.

We next determined the unresolved CXB flux for ACIS and EPIC observations (see Appendix \ref{CXB_flux}). We then created an X-ray background (XRB) image for each using the CXB flux and the galactic foreground flux from Table \ref{tab:galactic_fits}. For XMM, we also created a soft proton background image using methods discussed in Appendix \ref{soft_proton}. All these images were combined to produce the total scaled background image for XMM (NXB + X-ray background and foreground + soft proton) and for {\chandra} (NXB + X-ray background and foreground). These total scaled background images were used to obtain background-subtracted, exposure-corrected images and were also used to create annular regions for spectral extraction. The exposure-corrected, background-subtracted images for the full FOV of {\chandra} are shown in the left panel of Figure \ref{fig:fov}.

\begin{figure*}
    \centering
        \includegraphics[width=\columnwidth, keepaspectratio]{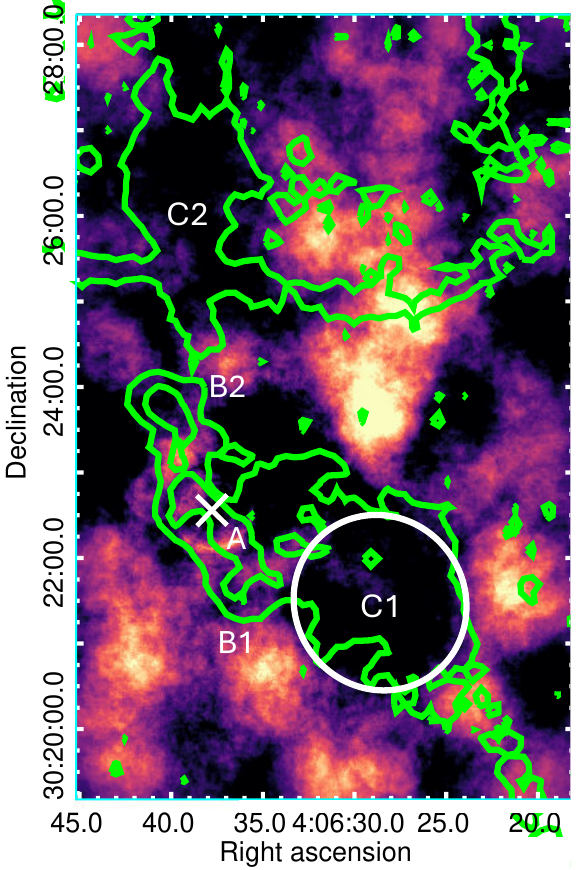}
        \includegraphics[width=\columnwidth, height=5.33in]{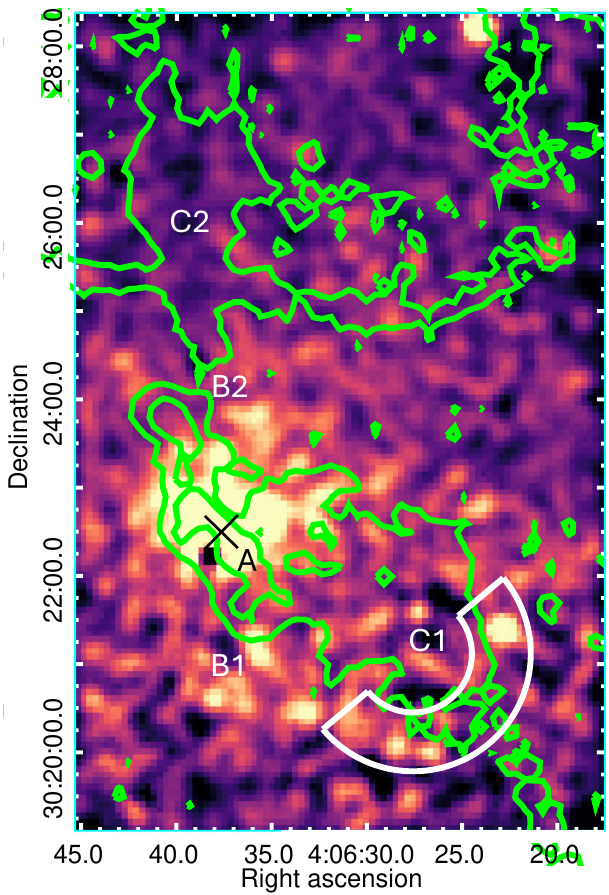}
    \caption{\small{\emph{Left:} {\chandra} residual map after subtracting a spherically symmetric double beta model fit to the surface brightness profile of Nest. A depression near the C1 lobe can be seen in the image and has been highlighted as a circular region. A further possible depression may be present near the C2 lobe. The image has been heavily Gaussian-smoothed with a $\sigma = 30''$ and $r = 30''$ for better visualization. \emph{Right:} {\xmm} exposure-corrected image showing the region around the C1 lobe. The image has been Gaussian-smoothed with a $\sigma = 6''$ and $r = 12''$ for better visualization. A possible limb-brightened feature around the C1 lobe is shown in the figure as a wedge. The LOFAR radio contours from Figure \ref{fig:fov} are shown in green. The X-ray center of the group has been marked with a cross in each figure. The different lobes have also been marked on each figure.}}
\label{NEST_chandra}
\end{figure*}

\section{Search for substructure in the X-ray images}  \label{lss}

We conducted a visual inspection of the X-ray images from {\chandra} and {\xmm} to identify any correlations or anti-correlations with the radio emission, which could indicate that the radio lobes are interacting with and potentially disturbing the IGrM. The images, on first look, appear smooth, with no visual evidence of discontinuities. We determined the center of the X-ray emission by weighting counts in the {\chandra} image within a one arcmin radius from the approximate center identified visually:

\begin{equation}
    \vec{R}_{\textrm{center}} (x,y) = \frac{\Sigma_i \vec{R_i} (x,y) c_i}{\Sigma_i c_i},
\end{equation}
where $c_i$ is the number of counts in the $i$-th pixel and $\vec{R_i}$ is a vector for the location of the pixel. Using this procedure, we obtained the X-ray center of Nest to be RA = 4:06:37.804, Dec = +30:22:35.351. This is in remarkable agreement (within an offset of $\sim 1$ arcsec) with the optical centroid of MCG+05-10-007 (RA=4:06:37.739, Dec=+30:22:34:769; \citealt{gaia20c}, \citealt{gaia21})\footnote{\href{https://tinyurl.com/3k26zwks}{SIMBAD}}.

To further probe whether any structures exist, we subtracted from the \textit{Chandra} image a fitted spherically symmetric double-beta surface brightness model. We show this subtracted image in the left panel of Figure \ref{NEST_chandra}. We used a double-beta profile instead of a single-beta profile to better model the steepening of surface brightness profile below 20 kpc (see Figure \ref{fig:SB} and \citealt{bri21} supplementary information). The two beta models had the same center and the fits were done using \texttt{Pyproffit v0.5}\footnote{\href{https://pyproffit.readthedocs.io/en/latest/index.html}{https://pyproffit.readthedocs.io/en/latest/index.html}} \citep{eck20}. The best-fit values are provided in Appendix \ref{sb_fits}, and fitted profiles are shown in Figure \ref{fig:SB}. Before fitting, all the point-source pixels in the {\chandra} counts image were replaced with counts interpolated from a defined background region. This region was chosen as an annular ellipse with the inner radius equal to the source radius and the outer radius 50\% larger. We used the \texttt{CIAO} tool \texttt{dmfilth} and \texttt{method=POISSON} to automate this task. 

As can be seen in Figure \ref{NEST_chandra}, no clear cavities are seen at the location of the A and B lobes with the present quality of data. However, we do identify a potential depression at the same location as the C1 radio bubble (Figure \ref{NEST_chandra} left panel). This suggests that the radio jet from the AGN is likely displacing the X-ray-emitting gas and creating a cavity. We also noticed a potential limb-brightened region in the XMM exposure-corrected image near the C-lobe upon zooming in (Figure \ref{NEST_chandra}, right panel).  This limb-brightened cavity was also detected by eROSITA \citep{bri21}. We did not create any residual image for XMM due to the presence of a chip gap near the region of the C1 bubble.

We measure the enthalpy of this cavity and the temperature of the limb-brightened region in Section \ref{cavity}. There is also a hint of another depression at the location of the C2 bubble in both \textit{Chandra} and \textit{XMM-Newton} image, although this is difficult to confirm with the present quality of data. Current data quality also did not visually reveal any features corresponding to the D lobes and D2 filament, although there is an indication of a density break in this region (See \S\ref{dens_break}). Any X-ray features that may be present in this region can help us to understand whether the filaments were created through a shock or as a consequence of the bubble evolution. 

\section{Spectral analysis} \label{spec_analysis}

\subsection{Region selection} \label{regions}

We divided the fields of view of {\chandra} and {\xmm} into annular bins with a constant signal-to-noise ratio. For \textit{Chandra}, the net source count was chosen to be 1600 (SNR = 40). This SNR ensures a small statistical error (a few percent) on temperature profiles, while allowing enough spatial resolution to investigate whether the object has a cool core. This choice provides us with 11 bins with a bin width of $20-45$ arcsec ($7-16$ kpc) over the field of view, with the outermost bin being 160 arcsec wide. 

For XMM, we chose the annular regions such that each bin has a net source count of at least $3600$ (SNR = $60$). Due to XMM's higher sensitivity in the $0.6-4.0$ keV band, a higher SNR threshold was chosen to investigate any temperature structures with better precision. At off-axis angles ($> 8$ arcmin from the aimpoint; see Figure \ref{fig:soft_proton}), the source becomes faint due to the vignetting effect, and thus the soft protons become the dominant source of counts. Hence, we restricted the spectral extraction to 8 arcmin from the aimpoint. This choice resulted in 11 different annular bins with a bin width of $35-65$ arcsec ($13-24$ kpc).

\subsection{Spectral extraction and fitting} \label{spectral_fitting}

For {\xmm}, we used the SAS tool \texttt{evselect} to extract spectra for each region from the observation, PN OoT, and FWC event files. The tasks \texttt{rmfgen} and \texttt{arfgen} were used to extract the redistribution matrix file (RMF) and the ancillary response file (ARF) for each region. To properly weigh the ARF, a detector map was provided for the extraction of such response files and the parameter \texttt{extendedsource} in \texttt{arfgen} was set to \texttt{yes}. For {\chandra}, the spectra from the observation and ``stowed" event files, as well as RMF and ARF, were obtained with the \texttt{CIAO} tool \texttt{specextract}. 

For {\xmm}, the NXB spectrum in each region was modeled with a broken powerlaw and a set of delta and Gaussian functions. This helped us to fit the quiescent state and the fluorescent lines in the particle background spectrum. The X-ray background and foreground components obtained in \S\ref{sky_bkg} were used to model the soft-energy background emission. The normalization of each component was scaled to match the area of each annular extraction region. The PN OoT spectrum was fitted with a spline model with the particle background components held fixed. The soft proton contribution for each annulus was then determined (See appendix \ref{soft_proton}). While fitting the total spectra, all of the above components were held fixed. The IGrM was modeled using a single \texttt{hot(cie)} model in \texttt{SPEX}. In each annulus, spectra from MOS1, MOS2, and PN were fit simultaneously. The spectra were binned optimally in \texttt{SPEX} before fitting \citep{kaa16}. We used the C-statistic for minimization during fitting. The fit quality was generally found to be statistically acceptable. The C-statistic in the outermost bin of XMM was $424$, while the expected C-statistic was $365 \pm 27$. Any systematic uncertainty in the background estimate can contribute to a higher C-statistic due to the low flux in this bin. The C-statistic in the previous bin, by comparison, was $356$, while the expected C-statistic was $362 \pm 27$.  The metallicity in the innermost three bins was allowed to vary, while it was kept fixed to a value of $0.2$ in the other bins. This was done to ensure the metallicity value was not unconstrained during the fits. A value of $0.2$ is consistent with similar metallicity values often seen in the outskirts of galaxy groups and clusters \citep{mer17}. 

While fitting {\chandra} spectra, the NXB and XRB components were also held fixed, and the IGrM component in each region was modeled with a \texttt{hot(cie)} model. The spectra from all ObsIDs were fitted simultaneously. The C-statistic in the outermost bin was $420$, while the expected C-statistic was $360 \pm 30$. All the fits were visually checked to confirm the absence of any systematic trends in the spectral residuals. A small systematic variation of the NXB between the stowed event files and the observed data may be responsible for the slightly higher-than-expected C-statistic. For the same reason as {\xmm} fits, the metallicity was allowed to vary in the innermost bin, while it was kept fixed to $0.2$ in the other bins. {\chandra} could only constrain metallicity in one bin since it has a lower effective area in the soft band\footnote{\href{https://cxc.cfa.harvard.edu/ciao/why/acisqecontamN0010.html}{https://cxc.cfa.harvard.edu/ciao/why/acisqecontamN0010.html}}$^,$\footnote{\href{https://xmm-tools.cosmos.esa.int/external/xmm_user_support/documentation/uhb/effareaonaxis.html}{https://xmm-tools.cosmos.esa.int/external/xmm\_user\_support/\\documentation/uhb/effareaonaxis.html}}. This diminishes {\chandra}'s ability to constrain the Fe-L complex at a temperature of a few keV. 

\begin{figure*}
    \centering
    \includegraphics[width=7in, height=8.7in]{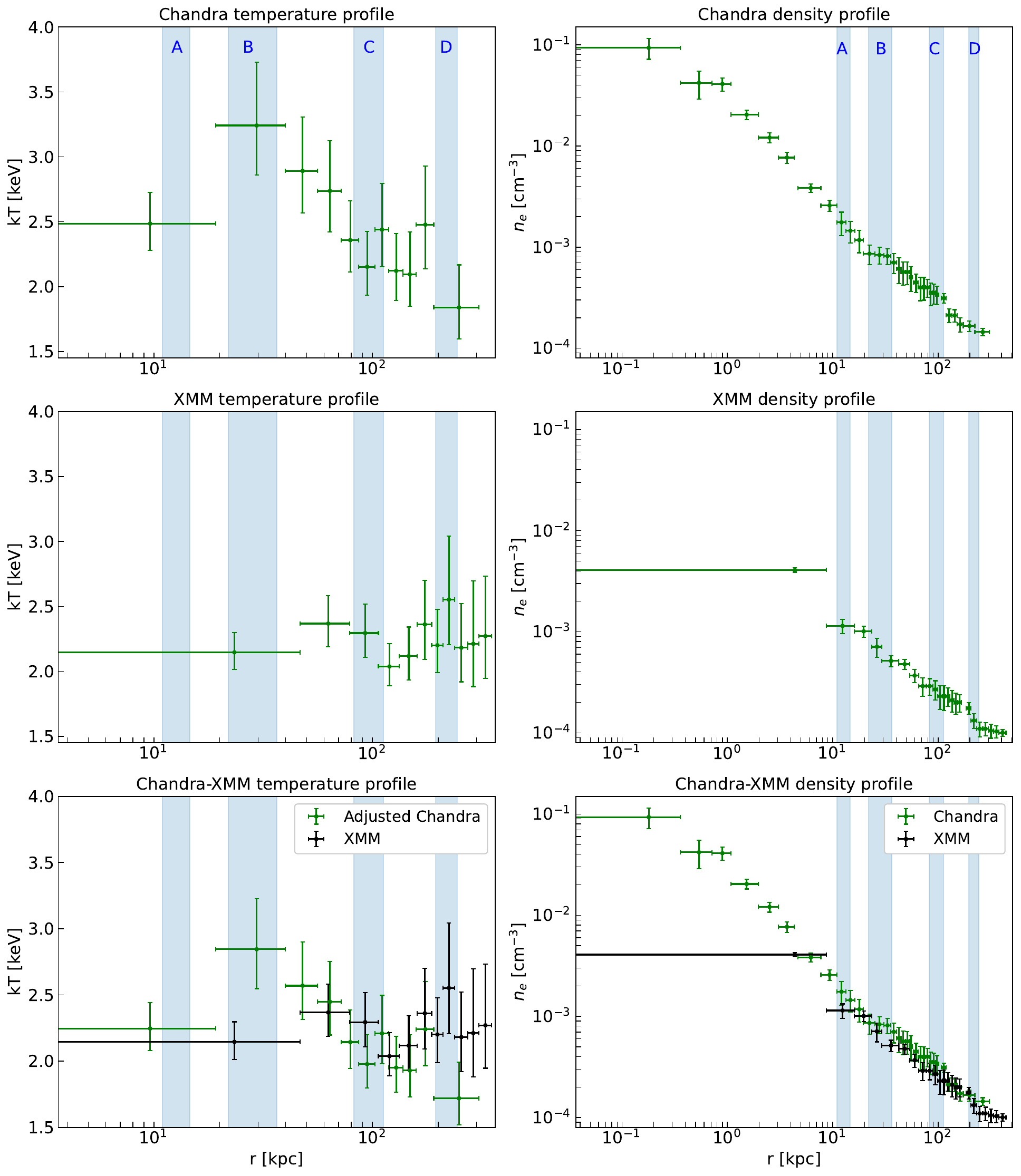}
    \caption{\small{Temperature profiles (left panel) and density profiles (right panel). \emph{First row:} {\chandra} temperature and deprojected density profile as a function of projected radius. \emph{Second row}: {\xmm} temperature and deprojected density profile as a function of projected radius. \emph{Third row}: Comparison between {\chandra} and {\xmm} temperature and density profiles after taking into account systematics. The shaded regions show the extent of the four lobes shown in Figures \ref{NEST_LOFAR} from the center of Nest.}}
    \label{fig:profiles}
\end{figure*}

\begin{figure}
    \centering
    \includegraphics[width=\columnwidth, keepaspectratio]{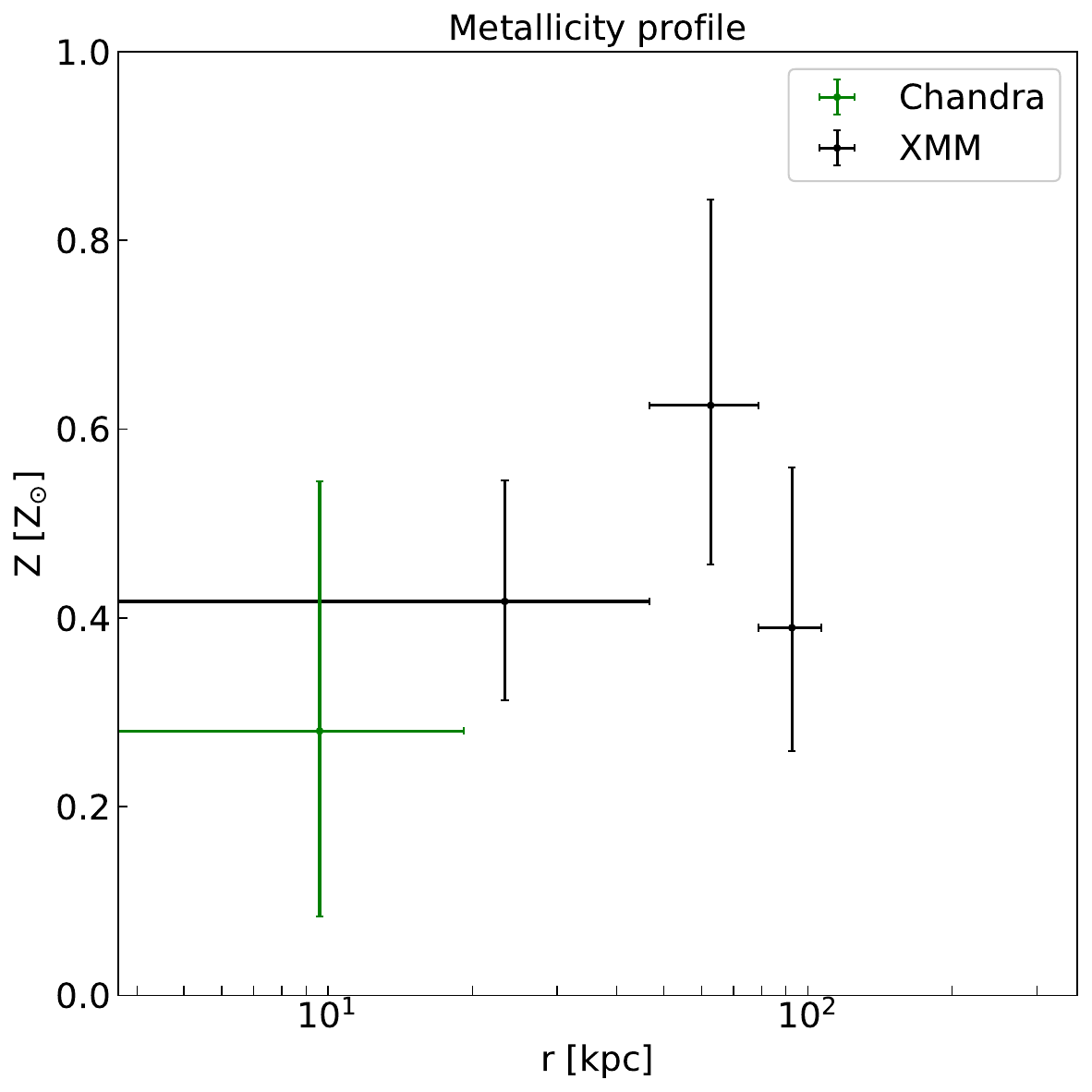}
    \caption{\small{Derived metallicity profiles from {\chandra} and {\xmm}. The green data point is the {\chandra} value and the black data points are {\xmm} values. Metallicity values beyond this region were assumed to be 0.2 solar during spectral fitting.}}
    \label{fig:metallicity_profiles}
\end{figure}

\begin{figure}
    \centering
    \includegraphics[width=\columnwidth, keepaspectratio]{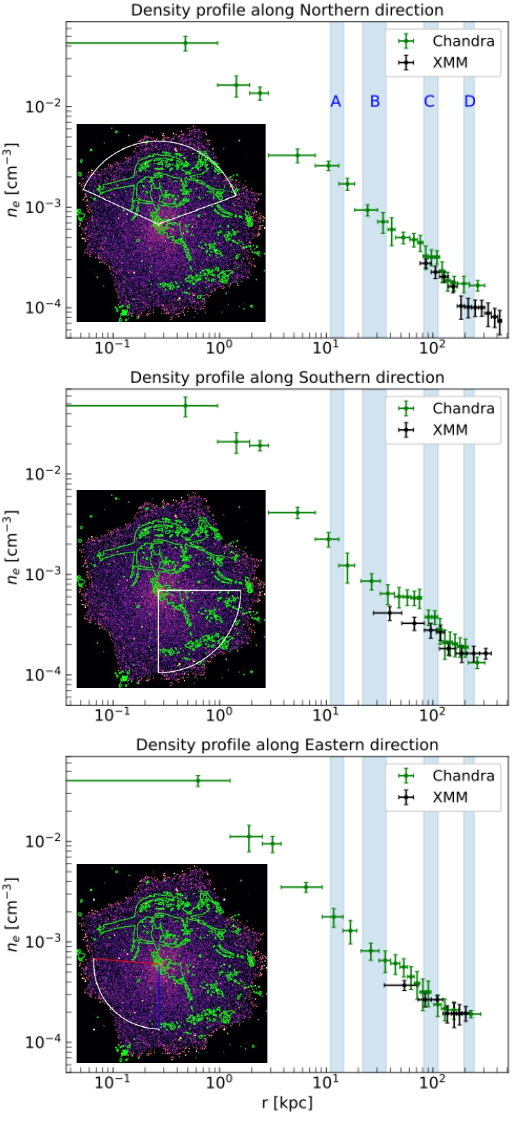}
    \caption{\small{\emph{Top:} Density profile along the northern lobes. \emph{Middle:} Density profile along the southern lobes. \emph{Bottom:} Density profile along the eastern direction where there is no radio emission. The shaded regions show the extent of the four lobes shown in Figures \ref{NEST_LOFAR} from the center of Nest. The inset figures show the extraction regions.}}
    \label{fig:direction_profiles}
\end{figure}

\section{Thermodynamic profiles} \label{profiles}

\subsection{Temperature and metallicity} \label{temperature}

The temperature profiles along the radial direction, using data from both telescopes, are shown in the left column of Figure \ref{fig:profiles}. We also show the outer edge of each pair of radio lobes in north and south directions with shaded regions. The robustness of the derived temperature values was tested by varying the normalization of the NXB, the galactic halo and the CXB. We find that the temperature value in the last bin is consistent within current statistical errorbars when the NXB normalization is varied up to 7\% and the galactic halo, CXB normalization is varied up to a factor of 2.  The systematic errors in our fits are thus smaller than the statistical errors.

The \textit{Chandra} temperature profile shows a tentative temperature drop ($\sim 0.8$ keV) below $30$ kpc ($80$ arcsec), suggesting the group might have a cool core. In addition, there are hints of a slight temperature increase around $100$ and $200$ kpc ($300$ and $500$ arcsec, respectively). This is also the distance of the C and D lobes from the center, as seen in the LOFAR images. The XMM measurements also suggest a temperature increase at a similar distance as the {\chandra} profile. However, the large errors in the temperature values prohibit us from concluding that there is a temperature jump at these locations. No temperature fluctuation is seen near the A and B lobes with the present quality of {\chandra} and {\xmm} data.

Finally, in the bottom left panel of Figure \ref{fig:profiles}, we show a comparison of the \textit{Chandra} and XMM temperature profiles after taking into account the systematic differences. The scaling relationship describing the offset between {\chandra} and {\xmm} temperatures can be written as \citep{sch15}

\begin{equation}
    \log_{10}\frac{kT_{I_{Y,\textrm{band}}}}{\textrm{1 keV}} = a \times \log_{10}\frac{kT_{I_{X,\textrm{band}}}}{\textrm{1 keV}} + b.
\end{equation}

\noindent We used the full band ($0.7-7.0$ keV) best-fit values for the comparison between Combined XMM and ACIS-I ($a = 0.889$, $b = 0$) as there is very little emission in Nest beyond $4$ keV. The error bars in the adjusted {\chandra} profile are larger than the original measurements because the statistical errors of the scaling parameters in \cite{sch15} were taken into account as well. The adjusted {\chandra} temperatures are in agreement with those from XMM, except at the core, where the larger XMM bin is expected to smooth the underlying temperature structures.

Next, we turn our attention to the metallicity profile. As noted in \S\ref{spectral_fitting}, the abundance could only be constrained in the innermost bin of \textit{Chandra} and the inner three bins of XMM. We show these derived metallicity values in Figure \ref{fig:metallicity_profiles}. As can be seen, XMM better constrains the abundance compared to \textit{Chandra}. The measurements suggest that there may be a drop in metallicity in the core of the group.

\subsection{Density} \label{density}

We also show deprojected density profiles from \textit{Chandra} and XMM in Figure \ref{fig:profiles}. The robustness of the derived density values was similarly checked as the temperature values. We again conclude that the effect of systematic uncertainty is lower than statistical uncertainty.

The deprojection of the density profile was done using the onion peeling method \citep{kris83} from surface brightness profiles in \texttt{PyProffit}. A constant temperature and metallicity of $2.1$ keV and $0.3$ $Z_{\odot}$ were assumed to obtain the count rate to the emissivity conversion factor. The temperature of $2.1$ keV corresponds to the XMM temperature in the innermost bin and is consistent with the temperature values throughout the field of view. We also found that the density values from \texttt{PyProffit} are mostly insensitive when the metallicity was varied between $0.2-0.4$ $Z_{\odot}$, i.e., the values within the central 20 kpc (see Figure \ref{fig:metallicity_profiles}). Any variation is within the current statistical error bars. Hence, we chose the median metallicity to obtain the density profiles. The spatial bins for density estimates were chosen such that each data point is at least $3\sigma$ significant. This criterion allows us to ensure that the errors are not too large due to the effects of onion deprojection, while at the same time ensuring a much more spatially resolved profile. We show this profile in the right column of Figure \ref{fig:profiles}.

The higher resolved density profiles suggest certain interesting features at the location of the radio lobes. In the {\chandra} profile, near the location of the C lobe, we see a possible change in the slope of the profile (see \S\ref{dens_break} for a quantitative analysis). In the XMM profile, we see a possible density break at the location of the D lobe. No change in slope or density jumps are detected near the location of the A lobes. Near the B lobe, a small slope discontinuity can be visually seen in the {\chandra} data, but no such feature is seen in XMM. This feature can either be spurious, or XMM's bigger PSF smooths out this feature. In the bottom panel, we show a direct comparison of the \textit{Chandra} and XMM profiles. We note that the {\chandra} profile reveals a steep density profile near the core with density values reaching 0.1 cm$^{-3}$.

The aforementioned features could be related to either the northern or southern lobes or both. To investigate this, we also extracted density profiles along various azimuthal sectors with the same $3\sigma$ SNR. These profiles are shown in Figure \ref{fig:direction_profiles}. We also show the extraction regions for these profiles in the inset of each sub-figure. As the XMM profile is smooth in the inner regions and affected by the broader PSF, we only show the XMM profiles in these directions near the location of the C and D lobes. 

We see that the jump near the D lobe is present in the northern direction, while the slope change near the C lobes is stronger in the southern direction (See \S\ref{dens_break} for fits). We note, though, that neither the \textit{Chandra} nor XMM fields of view extend significantly beyond the edge of the southern D lobe. We also extracted the density profile along the eastern direction where there is no radio emission to check whether the density jumps are indeed due to the radio lobes. As expected, we do not see any density jump in this direction. These results suggest that the AGN may be responsible for disrupting the IGrM and may also be causing the temperature increase near C and D lobe locations (as discussed in \S\ref{temperature}).

\subsection{Mass estimates} \label{mass_profile}

\cite{bri21} reported a total X-ray mass of $M_{500} = (3-7) \times 10^{13} M_{\odot}$ using mass-temperature scaling relations. Using the density and temperature profile reported before, we provide here an estimate for $M_{2500}$ and an updated $M_{500}$. Under the assumption of hydrostatic equilibrium, the total mass can be calculated as

\begin{equation}
    M(<r) = \frac{-kTr^2}{\mu m_p G}\Bigg(\frac{1}{n_e}\frac{dn_e}{dr} + \frac{1}{T}\frac{dT}{dr}\Bigg),
\end{equation}
where $\mu$ is reduced mass factor with a value of $0.596$, $m_p$ is the proton and $G$ is the gravitational constant. 

To compute the mass, we first fitted a parametric density and temperature model widely used in the literature \citep{vik06} to the derived profiles. Since the parameters in \cite{vik06} model are highly degenerate, we only vary the parameters that could be constrained by our temperature and density profiles (see \S\ref{vikhlinin_fits}). Using those parameters, we estimated $M_{2500} \sim 1.4 \times 10^{13} M_{\odot}$ and $M_{500} \sim 3 \times 10^{13} M_{\odot}$. The $M_{2500}$ and $M_{500}$ values correspond to $R_{2500} \sim 215$ kpc and $R_{500} \sim 480$ kpc, respectively. 

With these values, we can also calculate the gas mass fraction, $f_{g,500} = M_{\textrm{gas}}/M$:

\begin{equation}
    M_{\textrm{gas}} = \int 4\pi r^2 \rho_g dr,
\end{equation}

where the gas density $\rho_g = 2.3 \mu m_p n_p$ and $n_p = n_e/1.21$ is the proton number density. The factor of $2.3$ is needed to convert proton number density to total number density according to the adopted abundance table from \cite{lod09}. We calculated $f_{g,500} \sim 4\%$ at $R_{500}$ by extrapolating our profiles. This value is typical of galaxy groups that are found in the literature (e.g., \citealt{gas07}).

\subsubsection*{Independent mass estimate using eROSITA} \label{erosita_mass}

We do note that the estimate of $M_{500}$ discussed above was obtained by extrapolating our density and temperature profiles to $R_{500}$. This was necessary because $R_{500}$ of Nest is outside the field of view of both {\chandra} and {\xmm}. This may lead to sizable systematic uncertainties in estimating $M_{500}$. Hence, we also provide an independent estimate using the eROSITA data described in \cite{bri21} that has larger field of view compared to {\chandra} and {\xmm}.  The density profile was independently measured by eROSITA from the center to 12 arcmin. The density outside XMM field of view was then approximated by a single beta model \citep{cava76} with a norm $ \sim 5 \times 10^{-4}$ cm$^{-3}$, $r_c \sim 4.9$ arcmin and $\beta \sim 0.494$. With this density profile, and an assumed isothermal gas with an eROSITA measured temperature of $2.2$ keV,  we independently obtain $M_{500} \sim 7 \times 10^{13} M_{\odot}$ and $R_{500} \sim 640$ kpc. We note that, due to the fainter nature of the outskirts, systematic uncertainties on the level of the sky background can affect the outer slope of the eROSITA density profile, and hence the mass estimation. All mass estimates are reported in Table \ref{tab:mass_estimates}. 

\begin{table}
    \centering
\caption{Mass estimates of Nest from {\chandra}, {\xmm}, and eROSITA.}
\label{tab:mass_estimates}
    \begin{tabular}{ccccc}
    \hline
    \hline
         Telescope&  $M_{2500}$&$R_{2500}$& $M_{500}$ &$R_{500}$\\
         &  ($10^{13} M_{\odot}$) &(kpc)& ($10^{13} M_{\odot}$) &(kpc)\\
         \hline
         {\chandra} + XMM&  $1.4 \times 10^{13}$&$215$& $3 \times 10^{13}$&$480$\\
         eROSITA&  $-$&$-$& $7 \times 10^{13}$&$640$\\
    \hline
    \hline
    \end{tabular}
    \tablefoot{The uncertain $M_{500}$ mass and $R_{500}$ radius from {\chandra} and {\xmm} has been re-estimated independently with eROSITA.}
\end{table}

\subsection{Pressure}

\begin{figure}
    \centering
    \includegraphics[width=0.93\columnwidth, keepaspectratio]{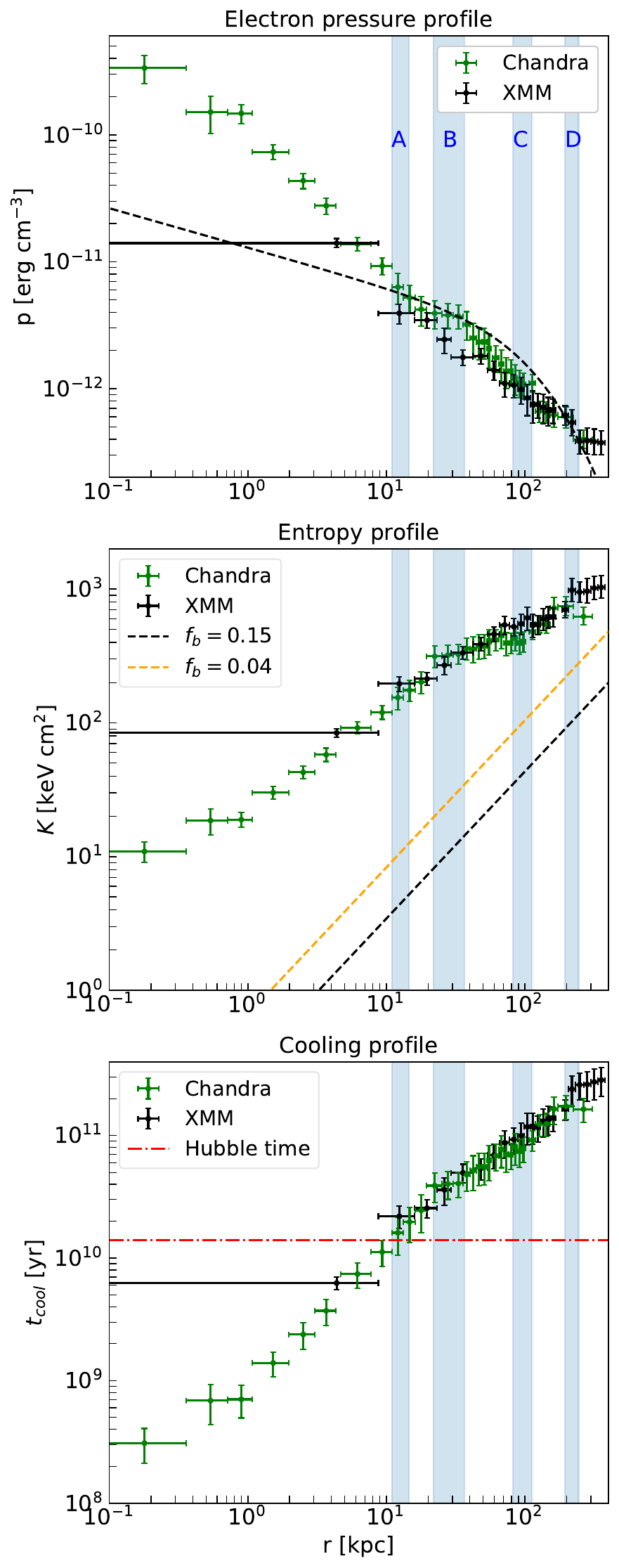}
    \caption{\small{\emph{Top:} Azimuthally averaged electron pressure profiles. The average expected profile from \textit{Planck} clusters is shown with a black dashed line. \emph{Middle:} Azimuthally averaged entropy profile. The self-similar $r^{1.1}$ profile for $f_b = 0.15$ is shown with a black dashed line, while the $f_b = 0.04$ profile is shown with an orange dashed line. \emph{Bottom:} Radial profile of the cooling time obtained from azimuthally averaged density, temperature, and metallicity profiles. The Hubble time $1/H_0$ is shown in red. The shaded regions show the extent of the four lobes shown in Figures \ref{NEST_LOFAR} from the center of Nest.}}
    \label{fig:other_profiles}
\end{figure}

We show azimuthally averaged pressure profiles from {\chandra} and XMM in Figure \ref{fig:other_profiles} (top panel). The electron pressure was calculated using the formula

\begin{equation}
    p = n_ekT,
\end{equation}
where $n_e$ is electron number density. We have corrected temperature systematics between \textit{Chandra} and XMM before overplotting the pressure profile. To calculate the pressure profile, we used the higher resolved bins of the density profile. As the temperature profile is less resolved, we used the nearest temperature values to calculate the pressure value in each bin. 

We note that similar to the density and temperature profiles, there are slope changes and jumps near the locations of the C and D lobes. We further compared our pressure profile with a generalized profile available in the literature \citep{nag07}:

\begin{equation}
    \frac{p(r)}{p_{500}} = \frac{p_0}{(c_{500}x)^{\gamma} [1 + (c_{500}x)^{\alpha}]^{(\beta-\alpha)/\gamma}},
\end{equation}

\noindent where, 

\begin{equation*}
    p_{500} = 1.45 \times 10^{-11} \textrm{ erg cm}^{-3} \Bigg(\frac{M_{500}}{10^{15}h^{-1}M_{\odot}}\Bigg)^{2/3} E(z)^{8/3}
\end{equation*}

\noindent and $x = r/r_{500}$.  We used the best-fit parameter values from 62 clusters in \cite{pla13}, [$p_0, c_{500}, \alpha, \beta, \gamma$] = [6.41, 1.81, 1.33, 4.13, 0.31],  to plot the black dashed line in the same figure. For our object, we obtain $P_{500} \sim 10^{-12}$ erg cm$^{-3}$. The Planck pressure profile is similar to the pressure profile in the earlier work of \cite{arn10} (See Figure 4 of \citealt{pla13} for a direct comparison). Although the generalized profiles are derived from clusters, \cite{sun11} showed that the median pressure profile of a sample of 43 nearby galaxy groups matches the \cite{arn10} universal profile well. Hence, in our work, we directly compare the pressure profile of our galaxy group object with the updated universal cluster profile from \cite{pla13}. 

The unusually steep density profile of our object may explain the deviation from the average expected profile at $r < 10$ kpc. We do, however, observe that the pressure in our object is consistently below the average profile from $10-200$ kpc. Beyond $200$ kpc, the average profile is below what we find in Nest. Such a characteristic possibly suggests that gas is being removed from the location of the radio lobes and ejected to larger distances. 

\subsection{Entropy}

The entropy profiles from {\chandra} and XMM are shown in Figure \ref{fig:other_profiles} (middle panel). The entropy was calculated using the formula

\begin{equation}
    K = kT/n_e^{2/3},
\end{equation}
where $n_e$ is the electron number density.

We also plot the expected entropy from purely gravitational heating. This self-similar entropy profile ($K_{SSC}$) was calculated following \cite{voi05} and \cite{pra10}:

\begin{equation}
    K_{SSC}(r) = 1.42K_{500}\Big(\frac{r}{R_{500}}\Big)^{1.1},
\end{equation}
with the self-similar normalization $K_{500}$ given by

\begin{equation}
    K_{500} = 106 \Bigg(\frac{M_{500}}{10^{14}M_{\odot}} \Bigg)^{2/3} f_b^{-2/3}E(z)^{-2/3},
\end{equation}
where $f_b$ is the baryon fraction and $E(z) = \sqrt{\Omega_m(1+z)^3 + \Omega_{\Lambda}}$. For  Nest, we obtain a $K_{500} = 171$ keV cm$^2$. We calculated the self-similar entropy profile using a universal baryon fraction value of 0.15 as well as our measured baryon fraction of 0.04. 

It is clear that the derived entropy is substantially higher than that from purely self-similar processes. Adopting a similar approach to \cite{eck25}, we used the expected self-similar entropy profiles to calculate the excess heat per particle ($\Delta Q$) in the IGrM by assuming an isochoric process (neglecting any cooling losses; \citealt{fin08,cha12}):

\begin{equation}
    \Delta Q  \approx \frac{kT}{\gamma - 1} \frac{K_{obs} - K_{SSC}}{K_{obs}},
\end{equation}
where $\gamma = 5/3$ is the adiabatic index and $K_{obs}$ is the observed entropy from data. This is a lower limit on the non-gravitational heat energy since gas ejected from the halo is not taken into account. The total heat energy could then be calculated as

\begin{equation}
    Q_{tot} (<R) = \int_0^R \frac{\Delta Q}{\mu m_p} 4\pi r^2 \rho_g  dr.
\end{equation}
Using the above formulas, we obtained the total excess heat over the whole field of view, $Q_{tot} \sim 6.5 \times 10^{60}$ erg for $f_b = 0.15$. For $f_b = 0.04$, we obtain $Q_{tot} \sim 5 \times 10^{60}$ erg. At least some of this excess heat is likely attributed to the central AGN activity, although the exact contribution is difficult to quantify since some entropy excess might have already been in place during the pre-heating phase (see Figure 9 of \citealt{mcc11}). We also note that the eROSITA mass estimate is higher (see \S\ref{erosita_mass}). If we use the eROSITA gas density and a baryon fraction of 0.15, we obtain $Q_{tot} \sim 6.1 \times 10^{60}$ erg.

\subsection{Cooling time}

The cooling time of a gas is defined as its thermal energy divided by the cooling rate. The cooling time, therefore, can be calculated as

\begin{equation}
    t_{cool} = \frac{p_{\textrm{tot}}}{(\gamma-1) n_en_H\Lambda(T,Z)},
\end{equation}
where $p_{\textrm{tot}} = 2.3p/1.21$ is the total gas pressure, $n_H \approx n_p = n_e/1.21$ is the hydrogen density, and $\Lambda(T,Z)$ is the cooling function. We calculated the cooling function at each spatial bin from the temperature, metallicity and density profile shown in Figure \ref{profiles}. The radial dependence of the cooling times is shown in Figure \ref{fig:other_profiles} (bottom panel) along with the Hubble time $1/H_0$.  It is clear that the gas below $10$ kpc may cool within the Hubble time. In the central region, the cooling time is as low as $\sim$$300$ Myr. However, both the radio morphology of the source and the radio spectral aging analysis \citep{bri25} suggest a very rapid duty cycle for the jet activity, with active phases of 50-100 Myr and very short inactive phases in between. The AGN can thus heat the IGrM much faster than the time the gas takes to cool.

\subsection{Binding energy versus excess heat}

To investigate excess heating compared to gravitational binding energy $E_{\textrm{bind}}$, we estimated $E_{\textrm{bind}}$ as

\begin{equation}
    E_{\textrm{bind}} (< R) = -G\int \frac{M(< r)}{r} 4\pi r^2 \rho_g dr. \label{binding_eq}
\end{equation}

\begin{figure}
    \centering
    \includegraphics[width=\columnwidth, keepaspectratio]{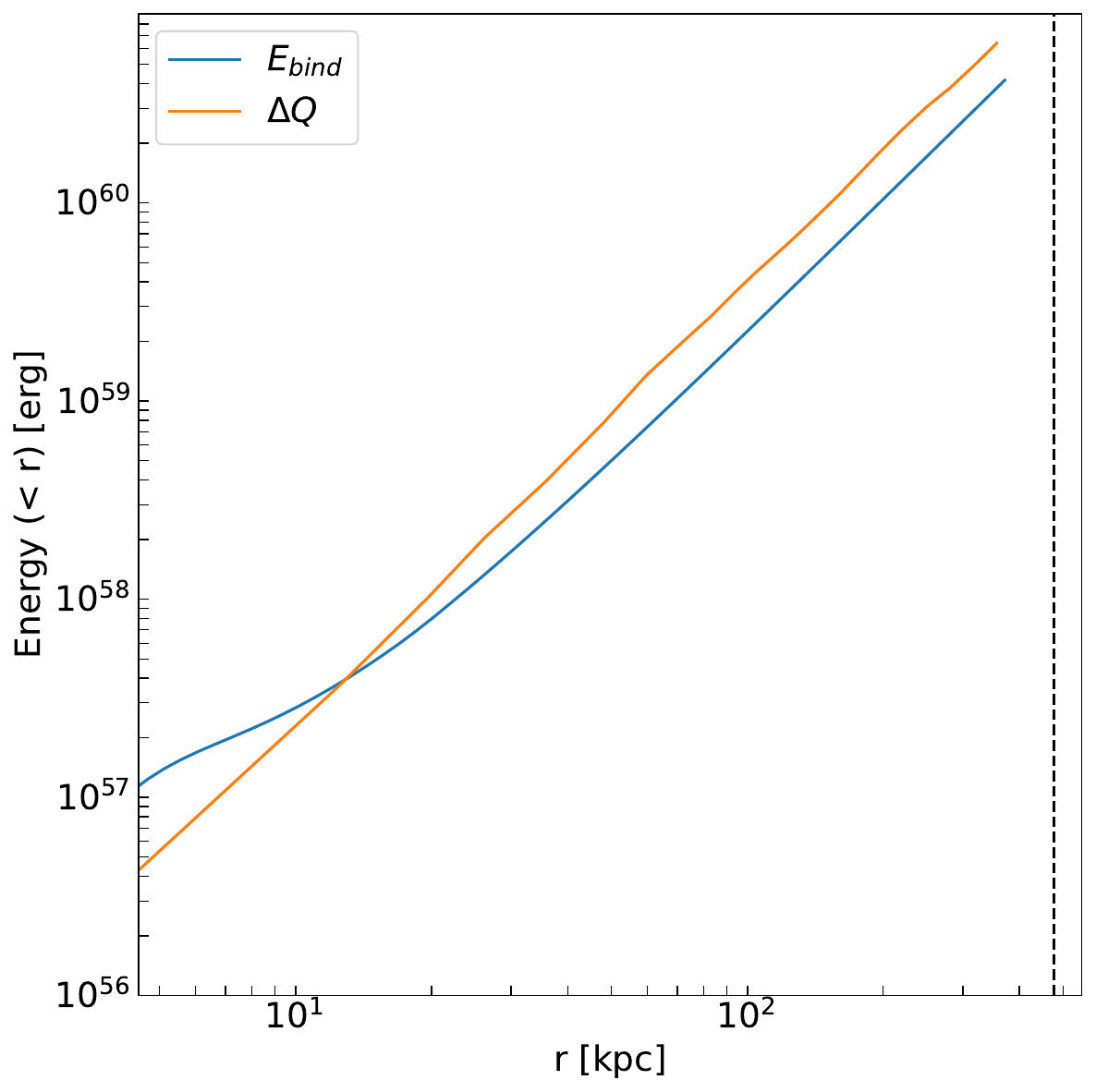}
    \caption{\small{Energy budget of the IGrM of the galaxy group Nest. The orange line shows the heating due to excess entropy. The blue line shows the binding energy of the object obtained through Equation \ref{binding_eq}. The black dashed line shows the location of $R_{500}$ from {\chandra} and {\xmm} data.}}
    \label{fig:energy_budget}
\end{figure}

Figure \ref{fig:energy_budget} shows the excess heat in the form of excess entropy and the binding energy as a function of radius. We used the {\chandra} and {\xmm} data up to the field of view to calculate these profiles. It is clear that in the inner regions of the group (< 13 kpc), the excess heat is lower than the binding energy. This further suggests that the heating is insufficient to prevent the formation of a cool core and is consistent with the temperature drop we obtained from the \textit{Chandra} profile. Beyond 13 kpc, the excess heat is higher, which may result in some gas being pushed out of the group towards the outskirts. 

\section{Density breaks} \label{dens_break}

In \S\ref{density}, we discussed a possible density slope change and density jump at the location of the C and D lobes. In this section, we aim to quantify the magnitude and significance of these breaks. In particular, we focus on two features: (1) the density slope change as seen in the {\chandra} data (Figure \ref{fig:direction_profiles}, middle panel) near the C lobe along the southern direction (2) the density jump as seen in the {\xmm} data (Figure \ref{fig:profiles}, middle right panel) near the D lobe in the azimuthally averaged profile. We chose the azimuthally averaged XMM profile instead of the northern or southern profile as they have insufficient counts near the D lobe to constrain the jump.  

\begin{figure}
    \centering
    \includegraphics[width=\columnwidth, keepaspectratio]{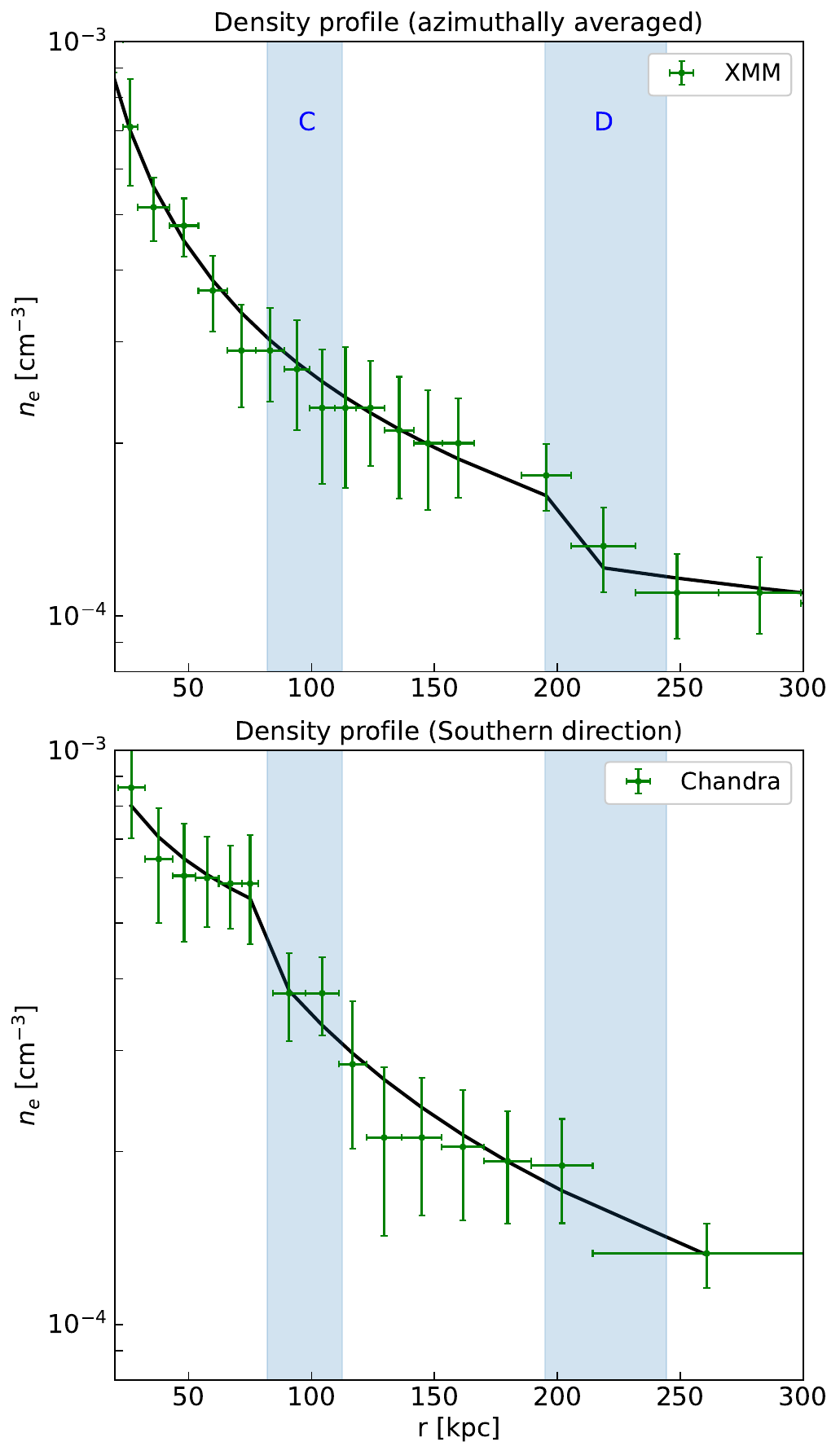}
    \caption{\small{\emph{Top:} Azimuthally averaged {\xmm} density profile. \emph{Bottom:} {\chandra} density profile along the southern lobes. The sector used to extract this profile is shown in Figure 6 inset. The black lines show the fitted broken powerlaw model. The shaded regions show the extent of the C and D lobes from the center of Nest.}}
    \label{fig:density_fits}
\end{figure}

We fitted a broken powerlaw model to these density profiles to constrain the slope change and jump. This model can be expressed as

\begin{equation*}
    n_e = n_1 \Big(\frac{r}{r_{br}}\Big)^{-\alpha_1} (r <= r_{br}) 
\end{equation*}
\begin{equation}
     n_e = n_2 \Big(\frac{r}{r_{br}}\Big)^{-\alpha_2} (r > r_{br})\ ,
\end{equation}
where $r_{br}$ is the break distance, $n_1$ is the density before the break and $n_2$ is density after the break. $\alpha_1$ and $\alpha_2$ are the slope of the density profile before and after the break, respectively. The fits are shown in Figure \ref{fig:density_fits}.

For the azimuthally averaged XMM density profile, at the edge of the D lobes, we obtained $n_1 = (1.55 \pm 0.15) \times 10^{-4}$ cm$^{-3}$, $n_2 = (1.21 \pm 0.14) \times 10^{-4}$ cm$^{-3}$. Defining the jump as $C = n_1/n_2$, we calculated a jump $C = 1.27 \pm 0.19$. This suggests a marginal $1.5\sigma$ deviation from a no-jump scenario ($C = 1$). If this jump is being created by a shock, the corresponding Mach number would be $\mathcal{M} = 1.19 \pm 0.14$. This can be obtained from $C$ by \citep{sar02}

\begin{equation}
  \frac{1}{C} = \frac{2}{\mathcal{M}^2(\gamma + 1)} + \frac{\gamma - 1}{\gamma + 1}.
\end{equation}
Such a Mach number is consistent with the values reported in the literature for objects with AGN-driven shocks \citep{for07,sim09,ran15,snio18}. We also calculated the energy, presuming that this jump is due to a shock. This energy can be calculated as \citep{dav01}

\begin{equation}
    E_S = \frac{1}{\gamma -1}(p_{2,\textrm{tot}} - p_{1,\textrm{tot}}) V_S,
\end{equation}
where $p_{2,\textrm{tot}}$ and $p_{1,\textrm{tot}}$ are pre- and post-shock total gas pressures and $V_S = 4/3 \pi r_{br}^3$ is the volume enclosed by the shock. We calculated the shock location to be $r_{br} = 218$ kpc, and with this value, we estimated an energy of $8.6 \times 10^{59}$ erg for the shock. This energy is consistent with that of shocks seen in group-scale objects (see \citealt{eck21} for a review). We note that the {\chandra} field of view does not extend much beyond 200 kpc. Hence, it is not possible to do a similar analysis using just {\chandra} data and compare that with the XMM analysis. In our calculation, we assume that the shock is spherically symmetric but note that this can be an overestimate if the shock propagates in a preferential direction. 

We do note that although the present evidence for a shock is weak, \cite{bri21,bri25} reported flattening of the radio spectral index at the location of the D2 filament with respect to the surrounding lobe emission, which might indicate re-acceleration or compression of particles $-$ possibly due to a weak shock produced by more recent AGN outbursts. From Figure \ref{fig:direction_profiles}, it is clear that the X-ray jump is primarily in the northern direction, i.e., at the location of the D lobes. The X-ray-detected jump may thus be associated with this speculated weak shock. It is not possible to determine the direction of the shock at this location with present data quality. 

For the {\chandra} density profile along the southern direction, we obtained $\alpha_1 = 0.4 \pm 0.2$ and $\alpha_2 = 1.0 \pm 0.2$. The uncertainties suggest that the significance of the slope change is $2.1\sigma$. We note that the XMM density profile does not show any slope change at the same location, as can be seen in Figure \ref{fig:direction_profiles}. There are two possibilities: (a) The C lobe and the C1 cavity unfortunately lie very close to an intersection of chip gaps in all three EPIC detectors. This causes a lack of sensitivity in this region. (b) This may simply be a spurious feature in the {\chandra} data. Higher-quality data may confirm or rule out the presence of these features in the future. (c) The larger PSF of XMM may smooth out the feature in this region. (d) It could be due to systematic uncertainties due to deprojection because, at larger radii, the azimuthal coverage of {\chandra} and XMM is different. We conclude that the present data quality provides low confidence in the presence of slope change and density jumps at the location of the C and D lobes.

\section{Cavity properties} \label{cavity}

\subsection{Cavity detection} \label{cavity_detection}

\begin{figure}
    \centering
    \includegraphics[width=\columnwidth]{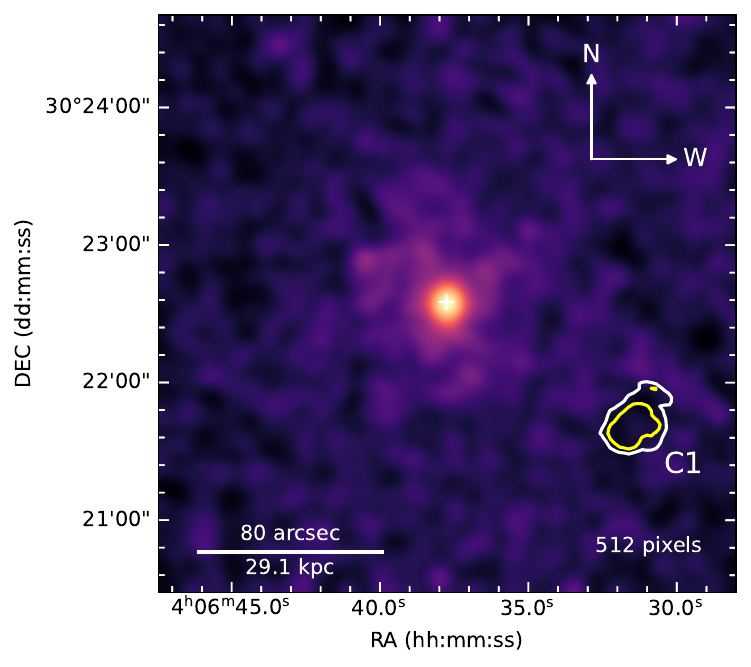}
    \caption{\small{Exposure-corrected {\chandra} image in the $0.5-7.0$ keV band overlaid with the contours of CADET prediction (white and yellow contours correspond to values of 0.4 and 0.6, respectively).}}
\label{cav_cadet}
\end{figure}

\subsubsection{CAvity DEtection Tool}

In Figure \ref{NEST_chandra}, we noticed a surface brightness depression in the {\chandra} residual image towards the southwest direction of the group. This was also detected by an analysis of the eROSITA data. To independently verify if this depression is significant, we applied the Cavity Detection Tool (CADET; \citealp{pls23}) to images from both {\chandra} and XMM. The CADET tool was developed primarily to detect cavities in {\chandra} images. We used the broad-band ($0.5-7.0$ keV) image and probed it on the following size scales: 128, 256, 384, 512 and 640 pixels (1 pixel = 0.492 arcsec). We used a broad-band image instead of the $0.6-4.0$ keV band because CADET was specifically trained to look for cavities in the broad-band, and the discrimination thresholds were tuned accordingly. Furthermore, our choice of scales was limited by the fact that the size of the CADET convolutional neural network was chosen to be 128 pixels. For \textit{XMM-Newton}, only scales of 128 and 256 pixels (1 pixel = 4 arcsec) could be probed as our choices were limited by the field of view. We furthermore chose the default discrimination threshold of 0.4 to look for cavities (see \citealt{pls23} for details). 

Based on the {\chandra} image, the CADET pipeline detected only a single X-ray cavity, which is co-spatial with the surface brightness depression C1 from Figure \ref{NEST_chandra}, and spatially co-aligned with the extended radio emission (Figure \ref{NEST_chandra}). We show this result in Figure \ref{cav_cadet}. When applied to an XMM image, the pixels in the CADET prediction at the position of the alleged X-ray cavity were also activated. Nevertheless, the prediction contained several other false positive cavities of similar significance, and thus, the detection based on XMM-Newton data is non-conclusive. We note, however, that the CADET pipeline has not been trained for XMM images. All three XMM instruments also have chip gaps near the location of the cavity. These reasons could cause CADET to be less reliable in the case of XMM.

\subsubsection{Azimuthal surface brightness analysis}

We also checked the significance of the X-ray cavity using azimuthal surface brightness profiles. We calculated the depth ($D$) and significance of the cavity similarly as in \cite{ube21}:
\begin{equation*}
    D = 1 - \frac{S_C}{S_M} 
\end{equation*}

\begin{equation}
    \text{significance} = \frac{D}{(1-D) \sqrt{\left( \frac{E_{\text{c}}}{S_{\text{c}}} \right)^2 + \left( \frac{E_{\text{M}}}{S_{\text{M}}} \right)^2}},
\end{equation}

\noindent where $S_{\text{c}}$ and $E_{\text{c}}$ represent the surface brightness and its uncertainty inside the C1 cavity region shown in Figure \ref{NEST_chandra}, $S_{\text{M}}$ and $E_{\text{M}}$ correspond to the median surface brightness and its uncertainty.  In case the cavity region extends over multiple angular bins, $S_C$ and $E_C$ are calculated by adding the counts in all such bins. The median surface brightness was calculated in an angular region that is three times the angular extent of the cavity, excluding the cavity region. The radial width was chosen so that it covers the entire radial extent of the cavity. According to our definition, the depth represents the fractional decrement of surface brightness inside of the cavity compared to the median surface brightness. 

\begin{figure}
    \centering
    \includegraphics[width=\columnwidth]{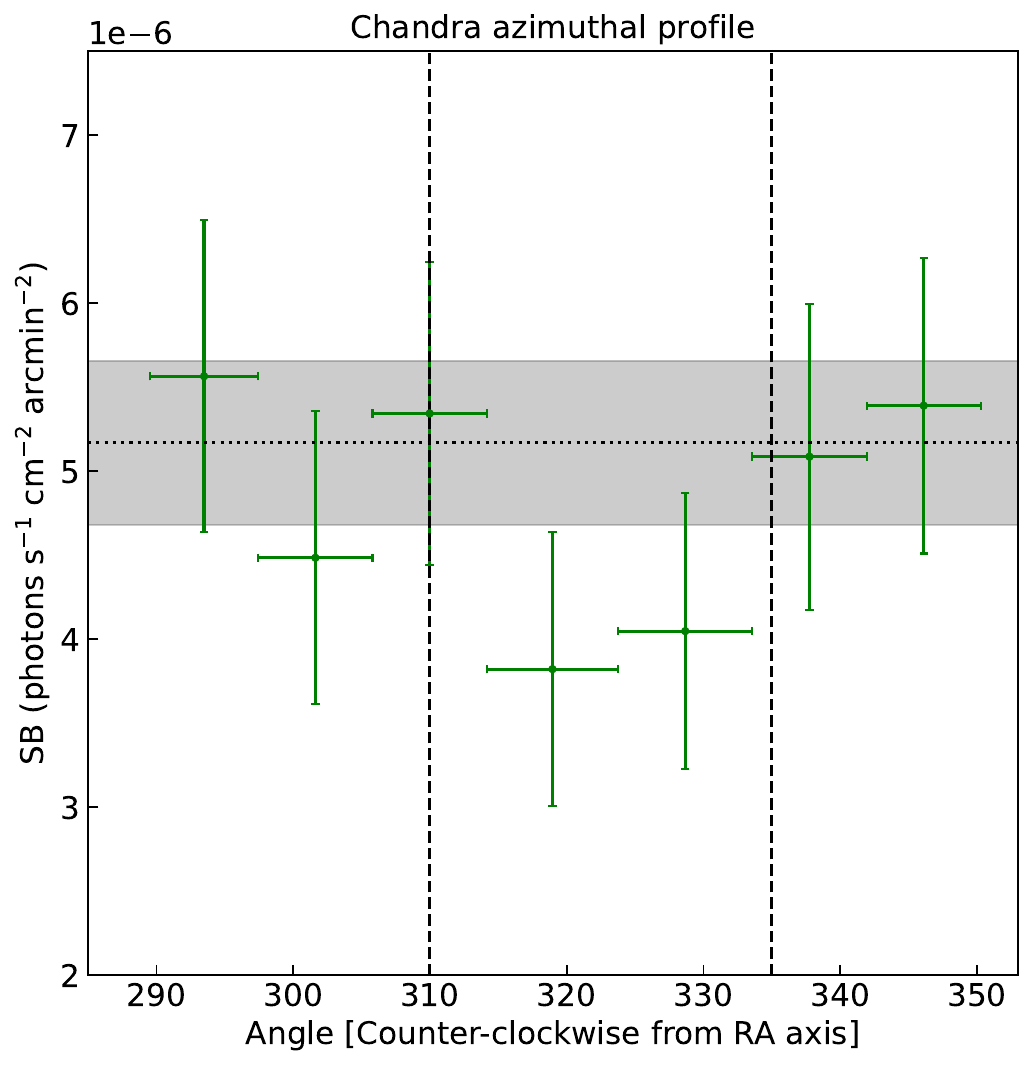}
    \caption{\small{{\chandra} azimuthal surface brightness profile (with SNR = 15) around the C1 cavity for $0.6-4.0$ keV image.  The extent of the cavity is shown with a black dashed line, while the median surface brightness is shown with a black dotted line. The error envelope for the median surface brightness is shown in black shaded region}}
\label{cav_significance}
\end{figure}

We calculated the azimuthal surface brightness profile using a constant SNR of 15 for {\chandra}. This ensured we obtained a few bins inside the cavity region. For this choice, the depth and the significance in the {\chandra} image were found to be $24.3\%$ and $2.6\sigma$, respectively. The azimuthal surface brightness profiles are shown in Figure \ref{cav_significance}. The XMM image, on the other hand, showed a much weaker significance ($\sim 1.6\sigma$) as all three detectors had chip gaps near the cavity. 

\subsection{Cavity energetics} \label{cav_energy}

To measure the thermodynamic properties of the cavity, we extracted spectra from the region shown in Figure \ref{NEST_chandra}. We measure the temperature and density from the region in the same way as the thermodynamic profiles. We measure a temperature of $1.8^{+0.3}_{-0.2}$ keV in the cavity region for \textit{Chandra} and $1.5^{+0.3}_{-0.2}$ keV for XMM. Assuming an ellipsoidal geometry for the cavity, the volume can be calculated as $V = 4\pi r_ar_br_c/3$, where $(r_a, r_b) = (22,23)$ kpc for our extraction region. Since we do not have information about the dimension of the cavity along line-of-sight, we assume the length along this dimension to be the same as $r_a$ (i.e., $r_c=r_a$). With this volume, the enthalpy can be calculated using the following relation (for $\gamma = 4/3$, i.e., a relativistic gas):

\begin{equation}
    H = 4p_{\textrm{tot}}V.
\end{equation}

The total gas pressure of the cavity was measured as $2.6^{+0.9}_{-0.7} \times 10^{-12}$ erg cm$^{-3}$ for \textit{Chandra} and $2.2^{+0.5}_{-0.4} \times 10^{-12}$ erg cm$^{-3}$ for XMM. From these pressure values, we obtained an enthalpy value of  $1.4^{+0.5}_{-0.4}  \times 10^{58}$ erg for \textit{Chandra} and $1.3^{+0.3}_{-0.2} \times 10^{58}$ erg for XMM. The cavity is located at a projected distance of $r_{\textrm{cav}} = 50$ kpc from the center, and thus the age of the cavity can be estimated as

\begin{equation}
    t_{\textrm{age}} = r_{\textrm{cav}}/c_s,
\end{equation}

\noindent where $c_s = \sqrt{\gamma kT/\mu m_p}$ is the local speed of sound. We calculated the sound speed and the corresponding sound travel time in each spatial bin using the temperature profile shown in Figure \ref{fig:profiles}. We added the sound travel time from each bin to obtain the age of the cavity, $t_{\textrm{age}}$. The sound speed in these bins varies between $\sim770-870$ km s$^{-1}$. We estimated $t_{\textrm{age}} = 59^{+5}_{-4}$ Myr for {\chandra} and $t_{\textrm{age}} = 64^{+2}_{-2}$ Myr for XMM. Based on this, we calculated the cavity power as

\begin{equation}
    P_{\textrm{cav}} = H/t_{\textrm{age}}.
\end{equation}

For {\chandra}, the resulting power of the cavity is $7^{+3}_{-2} \times 10^{42}$ erg s$^{-1}$, while for XMM, it is $6.0^{+1.5}_{-1.2} \times 10^{42}$ erg s$^{-1}$. We note that these values are upper limits since the distance to the cavity is measured in projected 2D space. We also used \texttt{SPEX} to calculate a luminosity of $8 \times 10^{41}$ erg s$^{-1}$ inside the cooling radius (r < 20 kpc), i.e., the radius within which the cooling time is less than the Hubble time. Although the cavity is now present outside the cooling radius, it might have deposited energy within the cooling radius during expansion. Assuming such a scenario, the cavity power will be more than enough to offset radiative losses. The cavity power and cooling luminosity for this galaxy group are consistent with the upper range of the scatter found in the literature (e.g., \citealt{eck21,hla22}). 

We further note that the cavity age was deduced higher in the \cite{bri21} analysis for the cavities at the location of C lobes ($t_{\textrm{age, Brienza}} = 130$ Myr). The higher age was calculated for a cavity at the location of C2 and using velocity dispersion of galaxies in the group \citep{tul15}. An updated analysis based on spectral ageing suggests that the age of the C1 cavity can be as high as $160$ Myr \citep{bri25}. \cite{bri21} also used an enthalpy of $p_{\textrm{tot}}V$ instead of $4p_{\textrm{tot}}V$ while estimating the power of the cavity. An enthalpy of $\sim p_{\textrm{tot}}V$ is suitable for calculating the amount of energy already deposited by the cavity due to expansion while rising through the IGrM to its current position. On the other hand, an enthalpy of $4p_{\textrm{tot}}V$ estimates the total energy that will be deposited by the cavity throughout the volume of the cluster. If we calculate the power of the cavity using the assumption of \citep{bri21}, i.e., $P_{\textrm{cav}} = p_{\textrm{tot}}V/t_{\textrm{age, Brienza}}$, we obtain  $P_{\textrm{cav}} \sim 6 \times 10^{41}$ erg s$^{-1}$. This is consistent with the $3 \times 10^{41} - 1 \times 10^{42}$ erg s$^{-1}$ power range quoted in \cite{bri21}.

We also constrained the temperature of the bright rim around the cavity, as seen in the {\xmm} image, to check if the expanding cavity is driving a shock into the IGrM. We find no discernible temperature jump that is expected from a shock. The temperature from {\chandra} was measured to be $2.0^{+0.3}_{-0.2}$ keV and that from XMM was measured to be $1.8^{+0.3}_{-0.2}$ keV. This measurement is consistent with previous estimates using eROSITA.

\section{The central supermassive black hole} \label{SMBH}

\subsection{Black hole mass}

\begin{figure}
    \centering
    \includegraphics[width=\columnwidth, keepaspectratio]{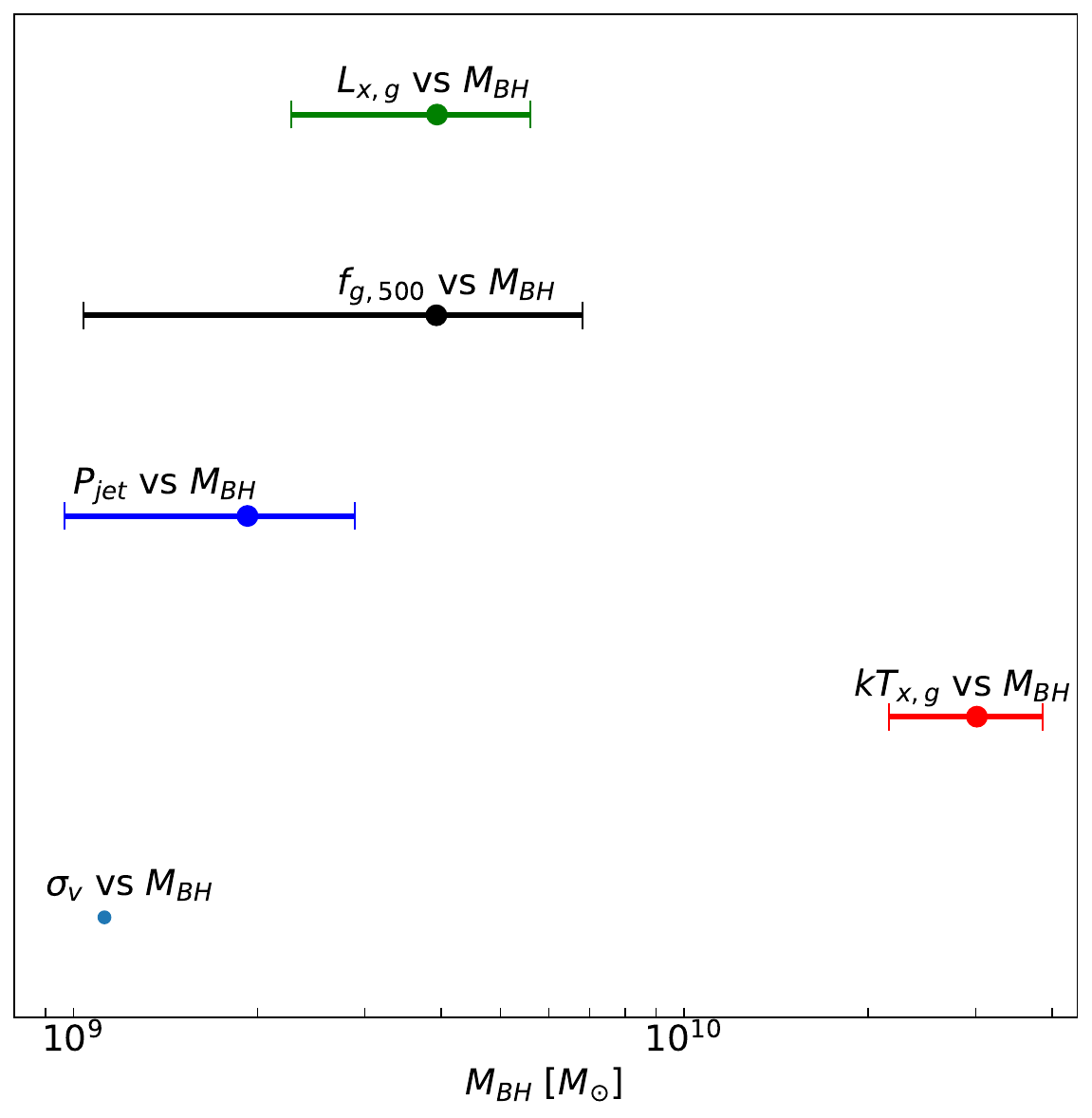}
    \caption{\small{Central black hole mass from scaling relations in Eqs. \ref{scaling_start}$-$\ref{scaling_end}. The mass of the black hole estimated from $L_{x,g}$ vs. $M_{\textrm{BH}}$, $M_{g,500}$ vs. $M_{\textrm{BH}}$, $P_{\textrm{cav}}$ vs. $M_{\textrm{BH}}$, $kT_{x,g}$ vs. $M_{\textrm{BH}}$, and $\sigma_v$ vs. $M_{\textrm{BH}}$ are $(4 \pm 2) \times 10^9 M_{\odot}$, $(4 \pm 3) \times 10^9 M_{\odot}$, $(1.9 \pm 0.9) \times 10^9 M_{\odot}$, $(30 \pm 9) \times 10^9 M_{\odot}$, and $1 \times 10^{9} M_{\odot}$, respectively.}}
    \label{fig:BH_mass}
\end{figure}

Using the temperature, cooling luminosity, gas mass fraction and power of the cavity, it is possible to obtain the central black hole mass, $M_{\textrm{BH}}$, by using the scaling relations reported in \cite{gas19} and \cite{pls22}. These relations can be written as

\begin{equation}
    \textrm{log}_{10}\Bigg(\frac{M_{\textrm{BH}}}{M_{\odot}} \Bigg) = (9.39 \pm 0.05) + (2.70 \pm 0.17)\textrm{ log}_{10}\Bigg(\frac{kT_{x,g}}{1 \textrm{ keV}} \Bigg) \label{scaling_start}
\end{equation}

\begin{equation}
    \textrm{log}_{10}\Bigg(\frac{M_{\textrm{BH}}}{M_{\odot}} \Bigg) = (10.63 \pm 0.15) + (0.51 \pm 0.04)\textrm{ log}_{10}\Bigg(\frac{L_{x,g}}{10^{44} \textrm{ erg s}^{-1}} \Bigg)
\end{equation}

\begin{equation}
    \textrm{log}_{10}\Bigg(\frac{M_{\textrm{BH}}}{M_{\odot}} \Bigg) = (11.02 \pm 0.24) + (1.11 \pm 0.11)\textrm{ log}_{10}(f_{g,500})
\end{equation}

\begin{equation}
    \textrm{log}_{10}\Bigg(\frac{P_{\textrm{jet}}}{10^{43} \textrm{erg s}^{-1}} \Bigg) = (-0.62 \pm 0.14) + (1.79 \pm 0.36)\textrm{ log}_{10}\Bigg(\frac{M_{\textrm{BH}}}{10^9M_{\odot}} \Bigg),
    \label{scaling_end}
\end{equation}

\noindent where $kT_{x,g}$ and $L_{x,g}$ are X-ray gas temperature and X-ray luminosity of the gas within $0.03 R_{500} \sim 14$ kpc. We assume $P_{\textrm{jet}} = P_{\textrm{cav}}$ for simplicity but note that $P_{\textrm{cav}}$ may be an underestimate if a cavity in the direction opposite to C1 also exists, i.e., at the location of C2. We report the results from these relations in Figure \ref{fig:BH_mass}. For black hole mass calculation, we used $kT_{x,g} = (2.5 \pm 0.2)$ keV, $L_{x,g} = (8.2 \pm 0.7) \times 10^{41}$ erg s$^{-1}$, $f_{g,500} \sim 0.04$ and $P_{\textrm{jet}} = 7^{+3}_{-2} \times 10^{42}$ erg s$^{-1}$. These values were obtained in the previous sections.

It is clear from the figure that the mass estimates from $L_{x,g}$ versus $M_{\textrm{BH}}$, $M_{g,500}$ versus $M_{\textrm{BH}}$,  and $P_{\textrm{jet}}$ versus $M_{\textrm{BH}}$ all agree with each other within $1.0-1.5\sigma$. As noted, the black hole mass from $P_{\textrm{jet}}$ is an underestimate if an X-ray cavity at the location of C2 also exists. However, the mass derived from $kT_{x,g}$ versus $M_{\textrm{BH}}$ disagrees with the others at $2.6-3.0 \sigma$ level. This is very likely because Nest is very hot near the center compared to the objects in the sample of \cite{gas19}. The hot core thus leads to an overestimate of the black hole mass and is an indication that the excess heat in Nest is high. 

It is also possible to estimate the black hole mass from the expected stellar velocity dispersion. The circular velocity can be estimated as

\begin{equation}
    v_{\textrm{circ}} = \sqrt{\frac{GM}{r}},
\end{equation}

\noindent where M is the mass and r is the distance from the center. Within central 10 kpc the gas pressure profile is well described by $p(r) \propto r
^{-1}$ (see top panel of Figure \ref{fig:other_profiles}) and the temperature $T \simeq$ constant $= 2.2$ keV (see top panel of Figure \ref{profiles}). Thus, the mass $M(r)$ is simply $kTr/(\mu m_pG)$ under the assumption of hydrostatic equilibrium. Then, $M \propto r$ and $v_{\textrm{circ}} =$ constant = 590 km s$^{-1}$ at r < 10 kpc where the central galaxy dominates the gravitational potential. The stellar velocity dispersion $\sigma_v$ profile thus strongly depends on a configuration of stellar orbits (e.g., \citealt{lok03}). Still, there exists a radial range where $\sigma_v$ only mildly depends on the anisotropy parameter which characterizes the orbital structure of a stellar system (see \citealt{chu10,lys14}). According to \cite{skr06}, the central galaxy in Nest200047 is well described by a Sersic surface brightness profile with $n=4$. For such an optical light profile the stellar velocity dispersion is weakly sensitive to the anisotropy parameter at $r=(0.5 - 1) R_{\textrm{eff}}$ (see Figure 3 in \citealt{chu10}), where $R_{\textrm{eff}}$ is the effective (half-light) radius of a galaxy and $R_{\textrm{eff}} \simeq 14''$ as measured in \cite{skr06}. At these radii, i.e. at $r \simeq 7 - 14$ arcsec,  $\sigma_v \simeq 0.6 v_{\textrm{circ}} \simeq 350$ km s$^{-1}$. The obtained velocity dispersion estimate is slightly higher than the central velocity dispersion $\sigma_c = 292.6 \pm 15.4$ km s$^{-1}$ measured within the inner $r \lesssim 2''$ \citep{van15}.

It is then possible to obtain a black hole mass using the well-known $M_{\textrm{BH}}-\sigma_v$ relation (e.g., \citealt{geb00,fer00,tre02}):

\begin{equation}
    \textrm{log}_{10}\Bigg(\frac{M_{\textrm{BH}}}{M_{\odot}}\Bigg) = \alpha + \beta\textrm{ log}_{10}\Bigg(\frac{\sigma_v}{200 \textrm{ km s}^{-1}}\Bigg). 
\end{equation}

\noindent This relation was calibrated using the velocity dispersion measured within a slit extending to the effective radius of a galaxy, so we used the obtained above estimate of the velocity dispersion at $r=(0.5 – 1) R_{\textrm{eff}}$ instead of the central value. For $(\alpha, \beta) = (8.13, 4.02)$ \citep{tre02}, we obtain $M_{\textrm{BH}} \sim 1 \times 10^9 M_{\odot}$. This value is consistent (within $1.0-1.5\sigma$) with estimates from $L_{x,g}$ versus $M_{\textrm{BH}}$, $M_{g,500}$ versus $M_{\textrm{BH}}$, and $P_{\textrm{cav}}$ versus $M_{\textrm{BH}}$ scaling relations. 

\subsection{Accretion power}

The central galaxy MCG+05-10-007 lacks any strong optical emission lines and is thus classified as a Low Excitation Radio Galaxy (LERG; see supplementary information in \citealt{bri21}). It has been argued in the literature that the LERGs accrete in a radiatively inefficient manner (e.g., \citealt{hard07,jan12}) that results from a geometrically thick flow, i.e., advection-dominated accretion flow  (ADAF; \citealt{nar95,abra95}). For such an accretion flow, the radiative efficiency of accretion can be very small ($\leq 1$ per cent), while the efficiency of the mechanical energy output remains high. Therefore, we first estimated the accretion efficiency of an idealized Bondi accretion \citep{bon52} into the central black hole using the following relation \citep{chu02}:

\begin{equation}
    s \approx 3.5 \Bigg( \frac{M_{\textrm{BH}}}{10^9 M_{\odot}} \Bigg)^{4/3} \left( \frac{\eta}{0.1} \right)^{2/3} \Bigg( \frac{L_X}{10^{43} \textrm{erg s}^{-1}} \Bigg)^{-2/3} \textrm{keV cm}^2,
\end{equation}

\noindent where $s$ is the entropy, $\eta$ is the accretion mechanical efficiency and $L_X$ is the cooling flow luminosity. This expression shows what gas entropy is needed to generate mechanical energy at a given rate via Bondi accretion. 
Using $L_X \sim 8 \times 10^{41}$ erg s$^{-1}$ (as reported in \S\ref{cav_energy}), the {\chandra} entropy value at the innermost radii ($s \sim 10.9$ keV cm$^2$) and $M_{\textrm{BH}} \sim 1\times 10^{9} M_{\odot}$, we obtain $\eta \sim 4.4 \times 10^{-2}$. The implicit assumption used here is that the gas radiative losses are compensated by the black hole the black hole output. The Bondi accretion rate is then

\begin{equation}
    \dot{M}_{B}=\frac{L_X}{\eta c^2}=1.4\times 10^{-3}\,M_\odot/\textrm{yr}.    
\end{equation}
This mass accretion rate is a lower limit since the entropy at the Bondi accretion radius can be lower than the entropy at the innermost {\chandra} data point. The Bondi accretion power will be the same as the X-ray cooling luminosity according to our assumptions. 

Observationally from the {\chandra} data, we find very little emission above 4.0 keV in the central region, suggesting that all of the observed emission is dominated by the diffuse thermal IGrM, with no evidence of non-thermal emission from a central point-like AGN in the X-rays. To estimate an upper limit to the central point-source flux, we fitted a single beta model to the surface brightness profile up to 10 arcsec ($\sim 4$ kpc) from the center in the $4.0-8.0$ keV band. The obtained fitted parameters were:  norm = $(4 \pm 3) \times 10^{-5}$ photons s$^{-1}$ cm$^{-2}$ arcmin$^{-2}$, $r_c = 2 \pm 2$ arcsec and $\beta = 0.5 \pm 0.3$. We obtained the flux of the central point source by extrapolating the single beta model to the center ($r = 0$) and subtracting the model value from the flux in the central $2$ pixel radius region (90\% of the {\chandra} PSF\footnote{\href{https://space.mit.edu/cxc/marx/tests/PSF.html}{https://space.mit.edu/cxc/marx/tests/PSF.html}}). From this calculation, we obtained a flux of $(2 \pm 4) \times 10^{-8}$ photons s$^{-1}$ cm$^{-2}$ or a luminosity of $(2 \pm 3) \times 10^{46}$ photons s$^{-1}$ in the $4-8$ keV band. For a powerlaw with a slope of $\Gamma = 1.7$ due to an X-ray corona in the central AGN, we calculate an upper limit on the $2-10$ keV luminosity of  $L_{\textrm{2-10 keV}} = (3 \pm 6) \times 10^{38}$ erg s$^{-1}$. Assuming a bolometric correction factor of $\sim$10 \citep{vasu07}, we estimated a $3\sigma$ upper limit of $2.1 \times 10^{40}$ erg s$^{-1}$ for the bolometric luminosity of the central black hole.

The above estimate suggests that the observed radiative power of the black hole is only $\sim$2.5\% of the Bondi accretion power. Radiative power being much smaller than the accretion power has been previously reported in the literature for multiple objects, such as IC4296 \citep{pell03}, M87 \citep{dim03}, Centaurus A \citep{evans04} and M84 \citep{bam23}. Since Nest contains prominent radio jets and little nuclear emission, almost all of the accretion power is being converted into mechanical energy. This is physically plausible since ADAF accretion is known to direct the gravitational potential energy of the black holes into jets rather than radiative output \citep{nar05,ho08}. We note that such an argument for low nuclear X-ray emission LERGs has been made previously in the literature (e.g., see \citealt{eva06,hard06,hard07})

\section{Summary and conclusions} \label{discussion}

In this work, we have used 140 ks of \textit{Chandra} and $\sim$25 ks of \textit{XMM-Newton} data to study the Nest200047 (hereafter Nest) galaxy group that was detected by LOFAR. We aimed to study the effects of AGN feedback on the IGrM and the properties of the central black hole. In this work, we demonstrated that:

\begin{itemize}
    \item The thermodynamic profiles in \S\ref{profiles} suggest that there is a cool core in Nest below $30$ kpc as indicated by the temperature profile.  The cooling profile further indicates that the cooling time at this distance is less than the Hubble time, and thus, the IGrM can form a cool core. Finally, the binding energy versus AGN heating profile suggests that the gravitational binding energy dominates the heating from the central AGN below $\sim 10$ kpc, and thus, the energy provided by the AGN will be insufficient to prevent the formation of a cool core. All of these pieces of evidence suggest that Nest likely has a cool core within $10-30$ kpc of the center.
    \item Upon extrapolating the thermodynamic profiles and assuming a hydrostatic profile, we obtain a mass of $(3-7)\times10^{13} M_{\odot}$ within $r_{500}$ using {\chandra}, {\xmm} and eROSITA. This is consistent with the previously estimated mass reported in \cite{bri21} using the mass-temperature scaling relation.
    \item Past radio jet activity from the central AGN likely inflated a cavity in the IGrM as it can be visually seen in unsharp-masked \textit{Chandra} images in \S\ref{lss}. This cavity was also visually seen in the eROSITA data. Independent analysis using the CADET pipeline and azimuthal surface brightness analysis in \S\ref{cavity_detection} further corroborates this suggestion. The azimuthal profile analysis indicates a cavity significance of $2.6\sigma$. The enthalpy of the cavity is estimated to be $\sim 1.4 \times 10^{58}$ erg and the power is $\sim 7 \times 10^{42}$ erg s$^{-1}$. We conclude that this one cavity has more than enough power to offset radiative losses since the cooling luminosity is only $\sim 10^{42}$ erg s$^{-1}$. However, more energy is likely being injected as there is evidence of multiple generations of bubbles in this system \citep{bri25}. 
    \item We observe some density breaks in the density profile along the southern direction as well as in the azimuthally averaged profiles. The statistical significance of these breaks is between $1.5-2.1\sigma$. We note that \cite{bri21,bri25} detected flattening of the radio spectral index at the location of the D2 filament. It was speculated to be due to re-acceleration or compression of particles due to a weak shock produced by more recent AGN outbursts. Assuming the detected break near the D lobes is due to the speculated shock, we estimated an energy injection of $8 \times 10^{59}$ erg in the IGrM assuming spherical geometry for the shock. If the shock exists, it would suggest that the dominant energy injection process is via shocks as opposed to energy deposition by cavities. 
    \item   There is significant excess entropy in Nest as evidenced by the entropy profiles in \S\ref{profiles}. The total excess energy is estimated to be $(5-6.5) \times 10^{60}$ erg depending on the assumed baryon fraction. The central AGN is likely responsible for some part of this energy. The energy from the central AGN is being added during an active phase of $50-100$ Myr \citep{bri25}.  
    \item The excess energy in the IGrM is higher than the binding energy above 13 kpc, suggesting that some gas can be pushed out of the object to the outskirts. There is some evidence of this in the pressure profiles in Figure \ref{fig:other_profiles}. The pressure profile of our object is consistently below the average observed cluster pressure profile between $10-200$ kpc. The total baryon fraction within $r_{500}$ is estimated to be $4\%$. This fraction is typical for galaxy groups \citep{gas07}.
    \item From scaling relations, we obtain a central black hole mass of $(1-4) \times 10^9 M_{\odot}$. The black hole mass derived from X-ray gas luminosity versus black hole mass,  baryon fraction versus black hole mass, jet power versus black hole mass, and velocity dispersion versus black hole mass relation are consistent with each other within $1.0-1.5\sigma$. However, the black hole mass obtained from X-ray gas temperature versus black hole mass relation is $2.6\sigma$ higher than that obtained from other relations. We argue that this is further evidence of the fact that the IGrM of Nest is super-heated, and this fact leads to a higher estimate of the black hole mass when using X-ray temperature.
    \item The AGN at the center of Nest is very faint with a bolometric luminosity $ < 2.1 \times 10^{40}$ erg s$^{-1}$ ($3\sigma$ upper limit). The Bondi accretion calculations suggest that the total accretion power should be much higher, i.e., $\sim 8 \times 10^{41}$ erg s$^{-1}$. Thus, only $\sim$ 2.5\% of the total accretion power is being channeled into radiative power. The rest of the energy is being channeled as jets instead. Such a scenario has been previously reported in the literature for accretion onto Low Excitation Radio Galaxies such as MCG+05-10-007 at the center of Nest200047.
\end{itemize}

The results of this work suggest that there is a large amount of excess heat in the IGrM of the Nest galaxy group, and the central AGN is likely responsible for some or most of this heating. In more massive objects (e.g., galaxy clusters), we know that the heating due to the AGN is typically a few times $10^{61}$ erg (e.g., \citealt{mcn05,wis07,maj24}). The upper limit on the AGN energy output in Nest ($6 \times 10^{60}$ erg) is thus comparable to that seen in much more massive haloes. Due to the enormous excess heat, Nest could be part of a class of overheated galaxy groups such as ESO 3060170 \citep{sun04}, AWM 4 \citep{gas08}, AWM 5 \citep{bald09} and SDSSTG 4436 \citep{eck25}, where there is significant excess entropy and heating. The high star formation quenching in these objects indicates that Nest could be suffering a similar fate. Follow-up studies with optical and infrared data will be needed to confirm such a hypothesis.

\section*{Data availability}
The \textit{Chandra} and \textit{XMM-Newton} data used in this article are publicly available. All the codes and intermediate data products used to generate the results of this paper are publicly available at \href{https://doi.org/10.5281/zenodo.15650741}{10.5281/zenodo.15650741}.

\begin{acknowledgements}
      This research has made use of data obtained from the \textit{Chandra} Data Archive provided by the \textit{Chandra} X-ray Center (CXC). We made use of \texttt{NUMPY} \citep{vdw11}, \texttt{SCIPY} \citep{jon01}, \texttt{ASTROPY} \citep{ast13}, and the \texttt{SBFIT} package\footnote{\href{https://github.com/xyzhang/sbfit}{https://github.com/xyzhang/sbfit}} \citep{zha21} in this work. Plots were generated using \texttt{MATPLOTLIB} \citep{hun07} and \texttt{APLPY}\footnote{\href{http://aplpy.github.com}{http://aplpy.github.com}}. AM thanks Elisa Costantini (SRON) and Liyi Gu (SRON) for discussions on the X-ray luminosity of the AGN at the center of Nest200047. TP thanks SRON for hosting his stay at the institute which enabled collaboration with AM and AS. The SRON Netherlands Institute for Space Research is supported financially by the Netherlands Organisation for Scientific Research (NWO). TP was supported by the GACR grant 21-13491X. M. Brienza acknowledges financial support from Next Generation EU funds within the National Recovery and Resilience Plan (PNRR), Mission 4 - Education and Research, Component 2 - From Research to Business (M4C2), Investment Line 3.1 - Strengthening and creation of Research Infrastructures, Project IR0000034 – “STILES - Strengthening the Italian Leadership in ELT and SKA”. M. Br\"uggen acknowledges funding by the Deutsche Forschungsgemeinschaft (DFG, German Research Foundation) under Germany's Excellence Strategy -- EXC 2121 ``Quantum Universe'' -- 390833306 and the DFG Research Group "Relativistic Jets". IK acknowledges support by the COMPLEX project from the European Research Council (ERC) under the European Union’s Horizon 2020 research and innovation program grant agreement ERC-2019-AdG 882679.

\end{acknowledgements}

\bibliographystyle{aa}
\bibliography{References} 

\begin{appendix}
\onecolumn

\section{Soft X-ray noise in the EMOS1 spectrum} \label{soft_noise}

\begin{figure*}[h!]
    \centering
    \includegraphics[width=0.75\columnwidth, keepaspectratio]{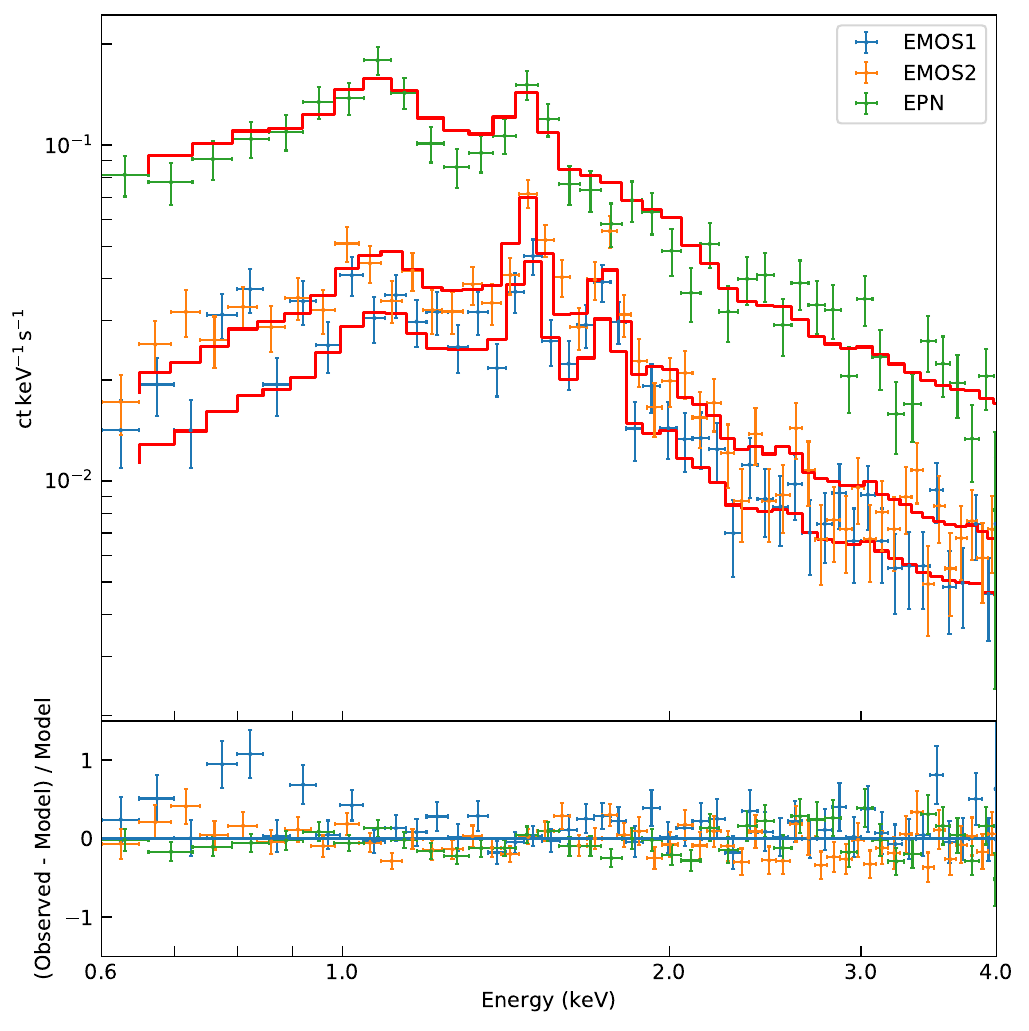}
    \caption{\small{MOS1 (blue), MOS2 (orange), and PN (green) spectra in the innermost XMM bin fitted simultaneously with \texttt{hot(cie)} + NXB + XRB + OoT spline model. During the fit the MOS1 spectrum below 1 keV was neglected.}}
    \label{fig:soft_noise_fig}
\end{figure*}

The fit for the innermost XMM bin is shown in Figure \ref{fig:soft_noise_fig}. We used a \texttt{hot(cie)} + NXB + XRB + OoT spline model to fit the spectrum from MOS1, MOS2, and PN simultaneously. To avoid soft noise in MOS1, we neglected the MOS1 spectrum below 1 keV. The \texttt{cstat} of the 156, while the expected \texttt{cstat} is $136 \pm 17$. Upon re-plotting the neglected part of the spectrum, we see that the MOS1 spectrum (blue) lies significantly above the model even though the MOS2 and PN spectrum agrees with the model. We speculate that this effect could be due to soft X-ray noise in the central CCD of MOS1 and is similar to what has already been reported in the literature for other MOS1 chips \citep{kun08}.

\section{Surface brightness fit parameters} \label{sb_fits}

The left panel of Figure \ref{NEST_chandra} was obtained by subtracting a spherically symmetric double beta model:

\begin{equation}
    I(r) = \frac{I_1}{(1 + r/r_{c1})^{\beta_1}} + \frac{I_2}{(1 + r/r_{c2})^{\beta_2}},
\end{equation}

\begin{figure*}[h!]
    \centering
    \includegraphics[width=0.59\columnwidth, keepaspectratio]{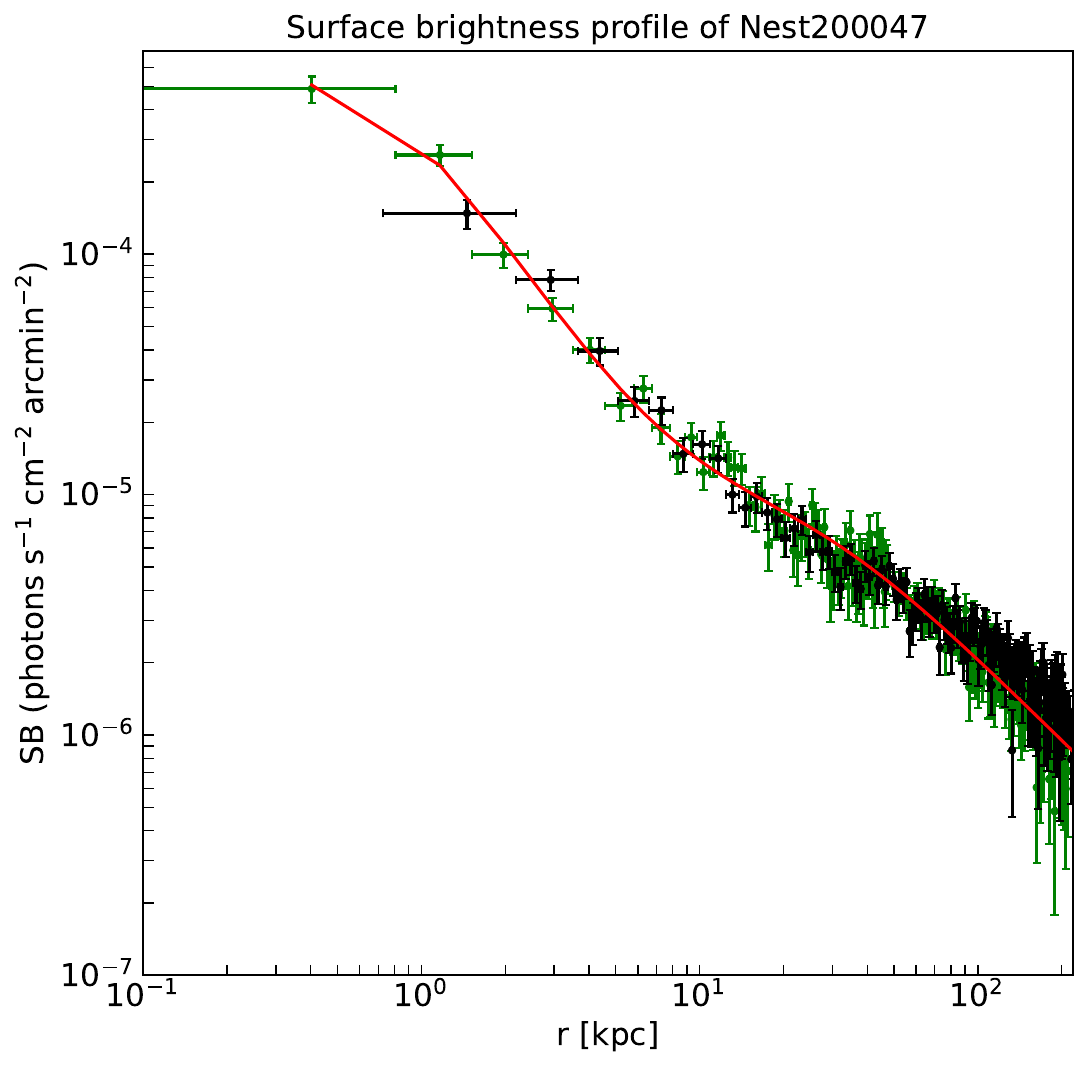}
    \caption{\small{{\chandra} surface brightness profile is shown in green, while the {\xmm} profile is shown in black. The fit to the {\chandra} profile used to obtain the left panel of Figure \ref{NEST_chandra} is shown in red.}}
    \label{fig:SB}
\end{figure*}

The fitted parameters, along with their errors, are shown in Table \ref{sb_param}. The {\chandra} and {\xmm} surface brightness profiles are shown in Figure \ref{fig:SB} along with the fit to the {\chandra} profile. Each data point in the profiles is at least $3\sigma$ significant.

\FloatBarrier

\section{CXB flux calculation} \label{CXB_flux}

The {\xmm} and {\chandra} observations of Nest has a deeper exposure compared to ROSAT all-sky data. Therefore, the CXB fluxes are lower as the telescopes can resolve more background point sources.  To accurately determine the CXB flux, we first calculated the flux of point sources detected with \texttt{wavdetect} assuming a local background and using an exposure map. The point source exclusion limit was then obtained by creating a histogram of point source fluxes and choosing the flux for which the number of point sources is highest. We obtained a flux limit of $5 \times 10^{-15}$ erg s$^{-1}$ cm$^{-2}$ in the $0.6-4.0$ keV band for the Nest XMM observation. Assuming a logN-logS relation identical to the {\chandra} Deep Field South \citep{leh12}, we obtained the unresolved CXB flux value of ($2.3 \pm 1.1$)$\times 10^{-15}$ erg s$^{-1}$ cm$^{-2}$ over a 1 arcmin$^2$ area in the $2-8$ keV range. This calculation was done using the publicly available software \texttt{cxbtools}\footnote{\href{https://github.com/jdeplaa/cxbtools?tab=readme-ov-file}{https://github.com/jdeplaa/cxbtools?tab=readme-ov-file}} \citep{mer15,deplaa17}. We find that this is $\sim$$54\%$ of {\rosat} CXB flux in the same energy range. In terms of the \texttt{pow} model (see Section \ref{sky_bkg}), a norm of $4.82 \times 10^{-7}$ photons/s/cm$^2$/keV for 1 arcmin$^2$ sky area is appropriate for our case and has been used in all subsequent XMM analysis.

Following a similar procedure for {\chandra}, we obtain a point source exclusion flux limit of $8 \times 10^{-16}$ erg s$^{-1}$ cm$^{-2}$ in the $0.6-4.0$ keV band. This suggests an unresolved CXB flux of ($1.2 \pm 0.3$) $\times 10^{-15}$ erg s$^{-1}$ cm$^{-2}$ over a 1 arcmin$^2$ area in the $2-8$ keV range which is $\sim$$30\%$ of {\rosat} CXB flux. Therefore, we used a \texttt{pow} norm of $2.63 \times 10^{-7}$ photons/s/cm$^2$/keV for 1 arcmin$^2$ sky area in all subsequent {\chandra} analysis.

\section{Soft proton calculation} \label{soft_proton}

\begin{figure*}[h!]
    \centering
    \includegraphics[width=0.52\columnwidth, keepaspectratio]{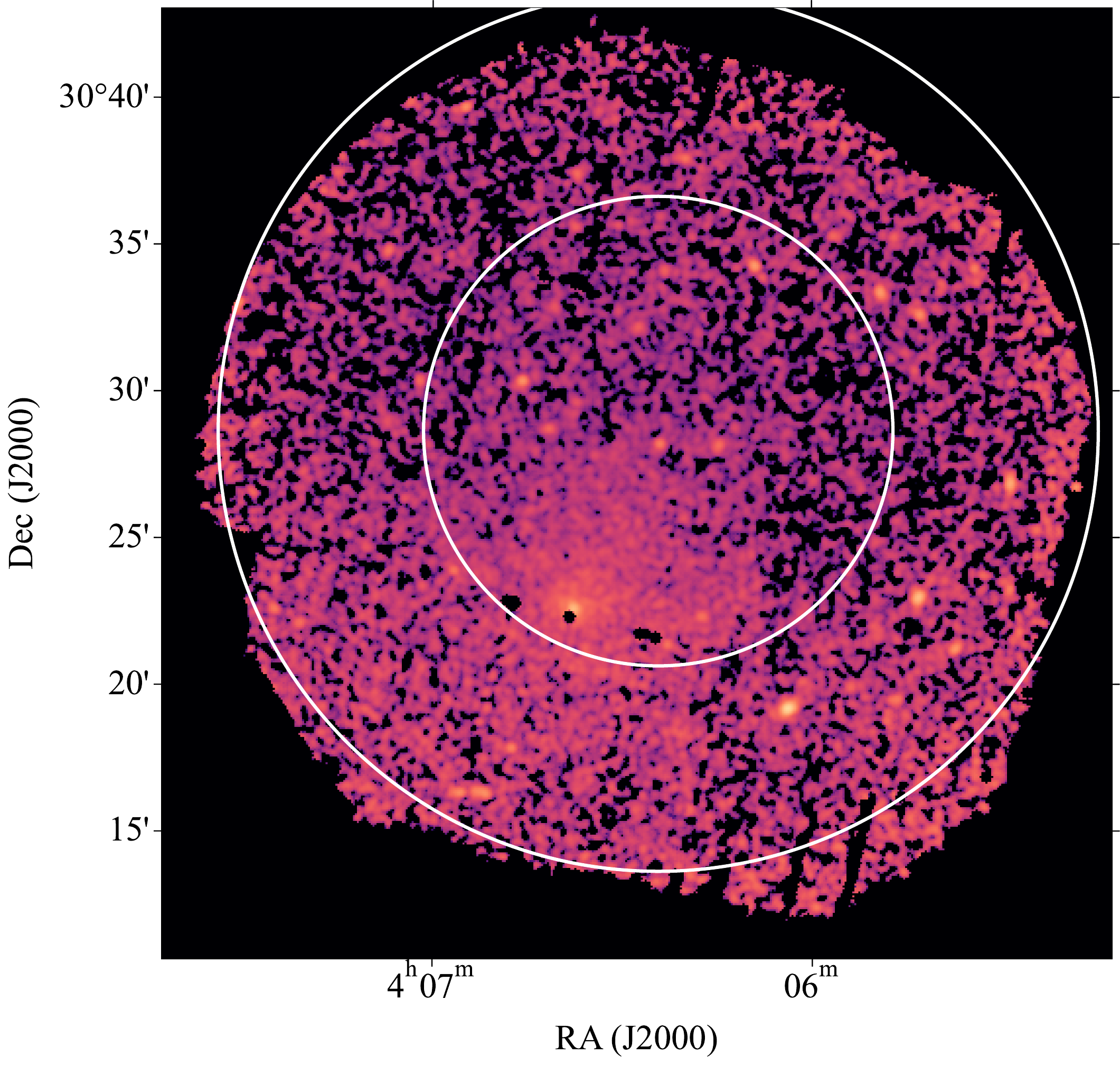}
    \caption{\small{Region used for analyzing and fitting soft proton. The region extends from 8 to 15 arcmin from the aimpoint.}}
    \label{fig:soft_proton}
\end{figure*}

As the XMM data is contaminated with soft protons at energies higher than a few keV, an accurate measure of soft proton flux was necessary to estimate IGrM temperatures.  We thus extracted spectra from all XMM instruments in the $8-15$ arcmin annulus centered at the aimpoint (see Figure \ref{fig:soft_proton}). These spectra were fit simultaneously in the $0.6-7.0$ keV range. This wider energy range was chosen, as opposed to $0.6-4.0$ keV, so that the soft proton contribution at higher energies is easier to constrain. We fit an absorbed \texttt{cie} to model the IGrM and a broken powerlaw to model the soft proton contribution. We ensured that the broken powerlaw model was not folded through the ARF. During the fit, non-X-ray background, PN OoT and X-ray background were held fixed. Among the broken powerlaw parameters, only the normalization was allowed to vary, while other parameter values were held fixed to the values obtained by \cite{zha20} (see Appendix B.1). We fixed the metallicity of the \texttt{cie} model to $0.2$ as well. The normalization of the soft proton for each instrument is reported in Table \ref{tab:sp_fits}.  The \texttt{cstat} for this fit was $320$, whereas the expected \texttt{cstat} was $260 \pm 20$.  The fit was visually checked to confirm the absence of any systematic residual. The soft proton contribution in any part of the CCD of any instrument can then be determined using the vignetting function described in Appendix B.2 of \cite{zha20}.

\FloatBarrier

\section{Density and temperature fit parameters} \label{vikhlinin_fits}

The fitted parameters to the density and temperature model described in \cite{vik06} are noted in Tables \ref{density_param} and \ref{temperature_param}. Only the values with errorbars could be constrained during the fit. 

\begin{table*}[h!]
\caption{Best-fit parameters for the surface brightness profile.}
\begin{tabular}{c c c c c c } 
\hline
\hline
 $I_1$& $r_{c1}$& $\beta_1$& $I_1$&    $r_{c1}$&$\beta_1$\\
            10$^{-6}$ photons s$^{-1}$ cm$^{-2}$ arcmin$^{-2}$  & (kpc)         &         &     10$^{-4}$ photons s$^{-1}$ cm$^{-2}$ arcmin$^{-2}$&       (kpc)         &   \\
\hline

 $8.7 \pm 0.6$&  $27.968 \pm 0.002$&   $0.355 \pm 0.006$&   $7.9 \pm 1.9$&   $0.69 \pm 0.16$& $0.47 \pm 0.02$\\
\hline
\end{tabular}
\label{sb_param}
\end{table*}

\vspace*{-\baselineskip}

\begin{table*}[h!]
    \centering
\caption{\texttt{SPEX} broken power-law normalization values for EMOS1, EMOS2, and EPN instruments.}
\label{tab:sp_fits}
    \begin{tabular}{cc}
    \hline
    \hline
         Instrument&  Flux ($0.6-7.0$ keV)\\
         &  (photons s$^{-1}$ cm$^{-2}$)\\
         \hline
         MOS1&  ($2.18 \pm 0.06$) $\times 10^{-1}$\\
         MOS2&  ($2.5 \pm 0.1$) $\times 10^{-1}$\\
 PN&($5.4 \pm 0.3$) $\times 10^{-1}$\\
 \hline
\hline
    \end{tabular}
    \tablefoot{The above parameters were constrained by fitting the model to a spectrum in an annular region extending from 8 arcmin to 15 arcmin.}
\end{table*}

\vspace*{-\baselineskip}

\begin{table*}[h!]
\caption{Best-fit parameters for density profile.}
\begin{tabular}{c c c c c c c c c c} 
\hline
\hline
 $n_0$                  & $r_c$         & $r_s$         & $\alpha$       & $\beta$                        & $\epsilon$                &    $\gamma$     &$n_{02}$               & $r_c$    &   $\beta_2$\\
            10$^{-3}$ cm$^{-3}$& (kpc)         & (kpc) &               &         &     &       &   10$^{-2}$ cm$^{-3}$      & (kpc) &     \\
\hline

 1.42&  8.2&  610&  $2.14 \pm 0.11$&   $0.242 \pm 0.004$&   $10^{-6}$&   3   & $2.5 \pm 0.2$&  6&  3.4\\
\hline
\end{tabular}
\tablefoot{Parameters that were held fixed while fitting are quoted without errors.}
\label{density_param}
\end{table*}

\vspace*{-1.75\baselineskip}

\begin{table*}[h!]
\caption{Parameters for the temperature profile.}
\begin{tabular}{c c c c c c c c } 
\hline
\hline
 $T_0$                  & $r_t$ & $a$ & $b$ & $c$ & $T_{min}/T_0$                  & $r_{cool}$             & $a_{cool}$    \\
            (keV)                  & (kpc) &     &     &     &                     & (kpc)                  & \\
\hline

  4.9& 18&   $10^{-6}$&  39&   $5 \times 10^{-2}$&  $6 \times 10^{-5}$&       55&       0.046\\
\hline
\end{tabular}
\tablefoot{Parameters that were held fixed while fitting are quoted without errors.}
\label{temperature_param}
\end{table*}

\end{appendix}
\end{document}